\documentclass[twocolumn,showpacs,amsmath,amssymb,superscriptaddress]{revtex4}
\usepackage{dcolumn}
\usepackage{bm}
\usepackage{amsmath}
\usepackage{amsfonts}
\usepackage{graphics}
\usepackage{graphicx}
\usepackage{color}

\begin{document}

\title{Quantitative law describing market dynamics before and after interest rate change}

\author{Alexander M. Petersen}
\affiliation{Center for Polymer Studies and Department of Physics, Boston University, Boston, Massachusetts 02215, USA}
\author{Fengzhong Wang}
\affiliation{Center for Polymer Studies and Department of Physics, Boston University, Boston, Massachusetts 02215, USA}
\author{Shlomo Havlin}
\affiliation{Center for Polymer Studies and Department of Physics, Boston University, Boston, Massachusetts 02215, USA}
\affiliation{Minerva Center and Department of Physics, Bar-Ilan University, Ramat-Gan 52900, Israel}
\author{H. Eugene Stanley}
\affiliation{Center for Polymer Studies and Department of Physics, Boston University, Boston, Massachusetts 02215, USA}

\date{\today}

\begin{abstract} 
We study the behavior of U.S. markets both before and after U.S.  Federal Open Market Committee (FOMC) 
meetings, and show that the announcement of a U.S. Federal Reserve rate change causes a financial shock, where the
dynamics  after the announcement is described by an analogue of the Omori earthquake law. We quantify the rate $n(t)$ of
aftershocks following an interest rate change at time $T$, and find power-law decay which scales as $n(t-T)\sim (t-T)^{-\Omega}$, with $\Omega$ positive. 
Surprisingly, we find that the same law describes the rate $n'(\vert t-T\vert)$ of ``pre-shocks" {\it before} the interest rate change at time $T$.
This is the first study to quantitatively relate the size of the  market response to the news which caused the shock and to uncover the presence of quantifiable preshocks. 
We demonstrate that the news associated with interest rate change is responsible for causing both the anticipation before the announcement and the surprise after the announcement.
We estimate the magnitude of financial news using the relative difference between the U. S. Treasury
Bill and the Federal Funds Effective rate.  Our results are consistent with the ``sign effect," in which ``bad news" has a larger impact
than ``good news." Furthermore, we observe significant volatility aftershocks, confirming a ``market underreaction" that  lasts at least 1 trading day.
\end{abstract}

\maketitle

\section{Introduction}
\label{sec:intro}
Interest rate changes by the Federal Reserve  provide a significant perturbation to financial markets, which
we analyze
 from the perspective of statistical physics \cite{Econophys0,Econophys1, Econophys2, Econophys3, 
Econophys4,Econophys5}. 
The Federal Reserve board (Fed), in charge of monetary policy as the central bank of the United States, is one of the
most influential financial institutions in the world. During Federal Open Market Committee (FOMC) meetings, 
the Fed determines whether or not to change key interest rates.  These interest rates serve as a benchmark and a
barometer for both American and international economies. The publicly released statements from the scheduled FOMC
meetings provide grounds for widespread speculation in financial markets, often with significant consequences. 

In this paper, we show that markets respond sharply to FOMC news in a complex way reminiscent of physical earthquakes
described by the Omori law \cite{Omori1, Omori2}. For financial markets, the  Omori law was first observed in market
crashes by Lillo and Mantegna
\cite{OmoriLillo}, followed by a further study of Weber {\it et al.} \cite{OmoriWeber}, which found the same behavior in
medium-sized aftershocks. However, the market crash is only an extreme example of information flow in financial markets. This
paper extends the Omori law observed in large financial crises to the more common Federal Reserve announcements, and
suggests that large market news dissipates via power-law relaxation (Omori law) of the volatility. 
In addition to the standard Omori dynamics following the announcement, 
we also find novel Omori dynamics before the announcement.

The  market dynamics following the release of FOMC news are consistent with previous  studies of price-discovery in foreign exchange markets following 
marcroeconomic news releases \cite{MacroNews1, MacroNews2}. Furthermore, we hypothesize that the uncertainty in Fed actions, coupled with 
the pre-announced schedule of FOMC meetings, can increase speculation among market traders, which can lead to the observed market underreaction \cite{sentiment}. Market
underreaction, 
meaning that markets take a finite time  to readjust prices following news, is 
 not consistent with the efficient market hypothesis; Several theories have been proposed to account for this phenomena \cite{noEMH}.

 We analyze all $66$ scheduled FOMC meetings in the eight-year period 2000-2008 using daily data from {\it
http://finance.yahoo.com}. Also, for the two-year period 2001-2002, we analyze the intraday behavior for 19 FOMC meetings using Trades And Quote (TAQ) data on the 1-minute time scale. 

The paper is organized as follows:
In Section \ref{sec:FOMC} we describe the FOMC meetings and the Fed interest rate relevant to our analysis. 
In Section \ref{sec:daily} we analyze the response of the S\&P100, the top 100 stocks (ranked by 12-month sales according to a 2002 {\it BusinessWeek} report) belonging to the 2002 S\&P 500 index, over
the 2000-2008 period using daily data. Using the relative spread between the 6-Month Treasury Bill and the Federal Funds
Effective rate, we relate the speculation {\it prior} to the FOMC meetings to the daily market volatility, measured here
as the logarithmic difference between the intraday high and low price for a given stock on the day of the announcement.
 In Section \ref{sec:Intraday} we study high-frequency intraday TAQ data on the 1-min scale for the S\&P100, and find  an Omori law with positive
exponent immediately following the announcement of Fed rate changes. Further, we relate the intraday market response,
(quantified by both the Omori exponent and Omori amplitude), to the change in market expectations before and after the announcement.

\section{FOMC Meetings, Fed Interest Rates and Treasury Bills}
\label{sec:FOMC} There are many economic indicators that determine the health of the U.S. economy. In turn, the health
of the U.S. economy sets a global standard due to the ubiquity of both the U.S. dollar and the economic presence
maintained through imports, exports, and the {\it Global} Market \cite{globalmarkets}. The U.S. Federal Reserve Target
rate, along with the Effective ``overnight" rate, set the scale for interest rates in the United States and abroad. The
Target rate is determined at FOMC meetings, which are scheduled throughout the year, with detailed minutes publicly
released from these meetings. The Effective rate is a  ``weighted average of rates on brokered trades''  between the Fed
and large banks and financial institutions, and is a market realization of the Target rate \cite{FED}.
 In Fig.~\ref{deltaexample} we plot the Federal interest rates
over the 8-year period 2000-2008.

Our analysis focuses on the FOMC meetings after January 2000. 
Historically, the methods for releasing the meeting details have varied. In the 1990s, there was a transition from a
very secretive policy towards the current transparent policy \cite{FOMCf}. 
 Since the year 2000, the Fed has released
statements detailing the views and goals of the FOMC. This increase in public information has led to an era of mass
speculation in the markets, revolving mainly around key economic indicators such as the unemployment rate, the Consumer
Price Index, the money supply, etc. These economic indicators also influence the FOMC in their decision to either
change or maintain key interest rates.  As a result of this open policy, the Fed has used the ``announcement effect" \cite{FOMCannouncementeffect}
to manipulate the federal funds market. Speculation has assumed many forms and new heights, evident in the implementation
of new types of derivatives based on federal securities. For instance, options and futures are available at the Chicago
Board of Trade which are based on Federal Funds, Treasury Bills, and Eurodollar foreign exchange. These contracts can be
used to estimate the implied probability of interest rate changes, utilizing sophisticated methods focussed on the price
movement of expiring derivative contracts \cite{FOMCnew, FOMCa, FOMCc, FOMCe, FOMCb, FOMCd}.

 In the next section, we outline a
simple method to measure speculation prior to a scheduled FOMC meeting using the 6-Month Treasury Bill and the Federal
Funds Effective (``overnight") rate.
 These data are readily available and are updated frequently at the website of the Federal Reserve \cite{FED}. 
Because each FOMC meeting is met with speculation (in the weeks before the meeting) and anticipation (in the hours
before the announcement), we identify the decision to change or not to change key interest rates as a market
perturbation. The market response results from the systematic stress associated with the  speculation and anticipation, 
which are not always in line with the FOMC decision. 

\section{Empirical Results}
\subsection{Response to FOMC Meetings on Daily Time Scale}
\label{sec:daily}
In this section we analyze the daily activity before and after 66 {\it scheduled } FOMC meetings  over the 8-year period
2000-2008, where scheduled meetings are publicly announced at least a year in advance \cite{FED}. 
We do not consider
unscheduled meetings resulting in rate change, which contain an intrinsic element of surprise, and are historically
infrequent (only 4 unexpected Target rate changes over the same period). Of primary importance, is the FOMC committee's
decision  to change or not change the Target rate $R(t)$ by some  percent $\Delta R(t)$, where the absolute
relative change $\vert \Delta R(t)/R(t-1) \vert$ has typically filled the range between $0.0$ and $0.25$. This section
serves as an initial motivation for the intraday analysis, and will also serve as a guide in developing a metric that
captures market speculation. 
In this section we use the intraday high-low price range to quantify the magnitude of price fluctuations.
 In particular, we analyze the companies belonging to the S\&P 100, and also the subset of 18 banking and finance
companies referred to here as the ``Bank'' sector.

In Fig.~\ref{deltaexample}(a) we plot $T(t)$, the time series for the 6-Month Treasury Bill, along with $F(t)$,
the {\it Federal Funds Effective
 rate}, and $R(t)$, the {\it Federal Funds Target rate},  over the 8-year period beginning in January 2000.
 The relative difference between the 6-Month Treasury Bill and the Federal Funds Effective
 rate is an indicator of the future expectations of the Federal Funds Target rate \cite{FOMCf}. Note that the 6-Month
Treasury Bill has anticipatory behavior with respect to the Federal Funds Target (and hence Effective) rates. Other more
sophisticated models utilize futures on Federal Funds and Eurodollar exchange, but these markets are rather new, and
represent the highly complex nature of contemporary markets and hedging programs \cite{FOMCnew, FOMCa, FOMCc, FOMCe, FOMCb,
FOMCd}. Hence, we use a simple and intuitive method for estimating market speculation and anticipation  by analyzing the
relative difference between the 6-Month Treasury Bill and the Federal Funds Effective rate. 

Fig.~\ref{deltaexample}(b) exhibits the typical interplay between the 6-Month T-Bill and the Federal Funds
Effective rate before and after a FOMC meeting. The change in the value of the Effective rate results from market
speculation, starting approximately one trading week (5 trading days) prior to the announcement. This change follows
from the forward-looking Treasury Bill, which in the example in Fig.~\ref{deltaexample}(b), is priced above the
Federal Funds rate even 15 trading days before the announcement. 

In order to quantify speculation and anticipation in the market prior to each scheduled FOMC meeting, we analyze the
time series $\delta(t)$ of the relative spread between $F(t)$ and  $T(t)$,
\begin{equation}
\delta(t) \equiv \ln \Bigl( \frac{F(t)}{T(t)} \Bigr) \ .
\label{deltat}
\end{equation}
As an example of this relation,  in Fig.~\ref{deltaexample}(c) we plot $\delta(t)$
for the 15 days before and after a typical FOMC meeting resulting in a rate change.
In order to study the speculation preceding the $i^{th}$ scheduled FOMC meeting, we calculate the average relative spread over the
 $L_{1}=15$ day period.
We weight the days in the $L_{1}$-day period leading up to the FOMC meeting day exponentially, such that the relative
spread on the $\Delta t^{th}$ day before the announcement has the weight $w(\Delta t) = e^{-\Delta t/\lambda}$. Without
loss of generality, we choose the value of $\lambda = 10$ days corresponding to two trading weeks \cite{ThetaDeltaPar}. 
We define the speculation metric,
\begin{eqnarray}
\Theta_{i} = \overline{\delta(t)}_{i} \equiv \frac{\sum_{\Delta t} \delta(t_{i}-\Delta t)w(\Delta t)}{\sum_{\Delta t}
w(\Delta t)} \ ,
\label{thetadef}
\end{eqnarray}
which is a weighted average of $\delta(t)$ before the announcement, where the sums are computed over the range $\Delta t
\in [1, L_{1}]$. The metric $\Theta_{i}$ for the $i^{th}$ FOMC meeting can be positive or negative, depending on the
market's forward-looking expectations.

In order to quantify the market response to the speculation $\Theta_{i}$, we analyze the market volatility around each
FOMC meeting.
 For a particular stock around the $i^{th}$ scheduled FOMC meeting, we take the daily high price $p_{\rm
hi}(t_{i}+\Delta t)$, and the daily low price $p_{\rm low}(t_{i}+\Delta t)$, for $\Delta t \in  [-20,20]$, where $\Delta
t=0$ corresponds to $t_{i}$, the day of the meeting. We then compute the high-low range  for each trading day,
\begin{equation}
r(t_{i}+\Delta t)\equiv \ln \Big (\frac{p_{\rm hi}(t_{i}+\Delta t)}{p_{\rm low}(t_{i}+\Delta t)}\Big ) \ .
\end{equation} 
For each stock and each meeting, we scale the range by $\langle r \rangle$, the average range over the 41-day time
sequence centered around the meeting day, resulting in the normalized volatility $v(t_{i}+\Delta t)\equiv r(t_{i}+\Delta
t)/\langle r \rangle$. Similarly, we use $\Phi(t_{i}+\Delta t)$, the time series for the volume traded over
the same period,
 to compute a weight for each stock corresponding to the normalized volume on the day of the FOMC meeting. We
calculate this weight as
 $\phi_{i} \equiv \Phi(t_{i})/\langle\Phi \rangle_{i}$, where $\langle\Phi \rangle_{i}$ is the average daily volume over
the 41-day time sequence centered around the $i^{th}$ meeting day. 
 We use a volume weight in order to emphasize the price-impact resulting from relatively high trading volume, since there are significant
 cross-correlations between volume change and price change \cite{VolPriceCC}. 
 Finally, we compute the weighted average volatility time series over all stocks and all meetings, 
  \begin{equation}
\langle v(\Delta t) \rangle \equiv \displaystyle \frac{\sum v(t_{i}+\Delta t) \phi_{i}}{\sum \phi_{i}} \ .
\label{eq1}
\end{equation}
In Fig. \ref{Daily} we plot the trend of average daily volatility defined in Eq.~(\ref{eq1}) for the 10 days before and 
after the scheduled announcements. 

We observe a peak  in  $\langle v(\Delta t) \rangle $ on FOMC meeting days, corresponding to 
$\Delta t=0$, with a more pronounced peak in the bank sector (Fig. \ref{Daily}). Stocks in the bank sector are
strongly impacted by changes in Fed rates, which immediately influence both their holding and lending rates. On average
there is a 15-20\% increase in volatility on days corresponding to FOMC meetings. 

In order to quantify  the impact of a single FOMC announcement on  day  $t_{i}$,  we define the average
market volatility
 \begin{equation}
 V_{i} = \langle v(t_{i}) \rangle \equiv \displaystyle \frac{\sum^{(i)}  v(t_{i}) \phi_{i}}{\sum^{(i)} \phi_{i}} \ .
\label{V0}
\end{equation}
Here, $\langle \dotsi \rangle_{i}$ and $\sum^{(i)}$ refer to the average and sum over records corresponding only to the
 day $t_{i}$. Again, $\phi_{i} \equiv \Phi(t_{i})/\langle\Phi \rangle$ is a
normalized weight, where now $\langle\Phi \rangle$ is the average daily volume over the entire 8-year period, since we compare many meetings across a large time span. 

In Fig.~\ref{ThetaDaily} we plot the average volatility  $V_{i}$ of the (a) S\&P100  and (b) the subset of 18 banking stocks versus $\Theta_{i}$. 
For negative values of $\Theta_{i}$, for which $T(t) > F(t)$ corresponding to an expected rate increase,
we observe a less volatile market response. Conversely, for larger positive values of $\Theta_{i}$,  for which $T(t) < F(t)$ corresponding to a rate cut,  there tends to be larger market fluctuations. Hence, the
market responds differently to falling and rising rates, where the direction in rate change often reflects the overall
health of the economy as viewed by the FOMC.  Typically, the FOMC implements rate increases to fight inflation, whereas rate decreases often 
follow bad economic news or economic emergency.
Hence,  our findings are consistent with the empirical ``sign effect", in which ``bad" news has a greater impact in markets than does ``good" news \cite{MacroNews2}. 
Furthermore, there is also a tendency for large average volatility even when
$\Theta_{i}$ is small, possibly stemming from the 
extreme surprise characteristic of some FOMC decisions. 
In these cases, more sophisticated methods are needed to improve the predictions of market movement.

\subsection{Intraday response to FOMC decision via an Omori Law}
\label{sec:Intraday} In the previous section we studied the market response on the daily scale. Now we ask the question,
``What is the intraday response to FOMC news?'' Here we analyze the TAQ data over the 2-year period Jan.~1, 2001 to
Dec.~31, 2002. The reported times for the FOMC announcement are listed in Table 1 \cite{data}. 
Inspired by the non-stationary nature of financial time series, methods have been developed within the framework of
non-equilibrium statistical mechanics to describe phenomena ranging from volatility clustering
\cite{ReturnIntervals1,ReturnIntervals3,ReturnIntervals2} to
 financial correlation matrices \cite{marketcorr0,marketcorr1, marketcorr2}. 
 
 We use the Omori law, originally proposed
in 1894 to describe the relaxation of after-shocks following earthquakes, to describe the response of the market to FOMC
announcements.  
Defined in Ref. \cite{OmoriLillo}, the Omori law quantifies the rate $n(t)$ of large volatility events
 following a singular perturbation at time $T$. The shock may be {\it exogenous} (resulting from external news stimuli)
or {\it endogenous} (resulting from internal correlations, e.g. ``herding effect'') \cite{ SornettePhysA,
SornetteEndoExo, information, earnings, SornettePNAS}. 
This rate is defined as,
\begin{equation}
n(\vert t-T \vert) \sim  \vert t-T \vert^{-\Omega} \ , 
\label{equation:rate}
\end{equation}
where $\Omega$ is the Omori power-law exponent.  

Here we study the rate of events greater than a volatility threshold
$q$, using the high-frequency intraday price time series $p(t)$. The intraday volatility (absolute returns) is expressed
as $v(t)
\equiv \vert \ln (p(t)/p(t-\delta t))\vert$, where we use $\delta t = 1$ minute. 
To compare stocks, we scale each raw time series in terms of the standard deviation over the entire period analyzed, and
then remove the average intraday trading pattern as described in Ref.~\cite{OmoriWeber}. 
This establishes a common volatility threshold $q$, in units of standard deviation, for all stocks analyzed.

In the analysis that follows, we focus on $N(\vert t-T \vert)$, the cumulative number of events above
threshold $q$,
\begin{equation}
N(\vert t-T \vert) =\int_{T}^{t}n(\vert t'-T \vert)dt' \equiv \beta (\vert t-T \vert)^{1-\Omega} \ ,
\label{equation:Ncum}
\end{equation}
which is less noisy compared to $n(\vert t-T \vert)$. Using $N(\vert t-T \vert)$, we examine the  intraday market
dynamics for  100 S\&P stocks, before $(t<T)$ and after $(t>T)$ the $i^{th}$ FOMC announcement at $T_{i}$, which 
typically occurs at 2:15 PM ET (285 minutes after the opening bell) for scheduled meetings. 

\begin{table}
\caption{ Reported times of market perturbations in the form of FOMC news. Dates of 19 FOMC meetings in the 2-year period
between  Jan. 2001 - Dec. 2002, where the Federal Funds Target rate $R_{new}$ 
was implemented by the rate change $\Delta R$ at $T$ minutes after the opening bell at 9:30 AM ET. The absolute
relative change $ \vert \frac{\Delta R}{R_{old}}\vert \equiv \vert \Delta R(t)/R(t-1) \vert$ has typically filled the
range between $0.0$ and $0.25$.
 Note: Date** refers to {\it unscheduled} meetings, in which the announcement time did not correspond to 2:15 PM ET ($T$
= 285 minutes)\cite{data}.}
\begin{tabular}{@{\vrule height 10.5pt depth4pt  width0pt}ccccc}\\
\noalign{
\vskip-11pt}
\vrule depth 6pt width 0pt \textbf{\em FOMC Date}  & \textbf{ $R_{new}$ (\%)}  & $\Delta R$ & $\frac{\Delta R}{R_{old}}$
& \textbf{\em T}  \\
\hline 01/03/01** & 6 & -0.5& -0.077 & 210 \\
01/31/01 & 5.5 & -0.5& -0.083 & 285\\
03/20/01 & 5 & -0.5& -0.091 & 285\\
04/18/01** & 4.5 & -0.5&-0.100 &  90\\
05/15/01 & 4 & -0.5& -0.111 & 285\\
06/27/01 & 3.75 & -0.25& -0.063& 285\\
08/21/01 & 3.5 & -0.25& -0.067& 285\\
09/17/01** & 3 & -0.5& -0.143& 0\\
10/02/01 & 2.5 & -0.5& -0.167& 285 \\
11/06/01 & 2 & -0.5 & -0.200 & 285\\
12/11/01 & 1.75 & -0.25& -0.125& 285\\
01/30/02 & 1.75 & 0 & 0.00& 285\\
03/19/02 & 1.75 & 0 & 0.00& 285\\
05/07/02 & 1.75 & 0 & 0.00& 285\\
06/26/02 & 1.75 & 0 & 0.00& 285\\
08/13/02 & 1.75 & 0 & 0.00& 285\\
09/24/02 & 1.75 & 0 & 0.00& 285\\
11/06/02 & 1.25 & -0.5 & -0.286& 285\\
12/10/02 & 1.25 & 0& 0.00& 285 \\
\hline
\end{tabular}
\label{table:dates0102}
\end{table}

In Figs. \ref{Omori}(a,b) we plot the average volatility response 
 $N(t)$ of the $S \equiv 100$  stocks analyzed, where
\begin{equation}
 N(t)  \equiv \frac{1}{S}\sum_{j=1}^{S} N^{j}(t) \ .
 \label{equation:AveNcum}
 \end{equation}
 This average is obtained by combining the individual Omori responses, $N^{j}(t)$, of the $S$ stocks. Such averaging does not cancel
the Omori law, but allows for better statistical regression. This is especially useful for an Omori law corresponding to
large volatility threshold $q$, where a single stock might not have a sufficient number of events. 
In Fig. \ref{Omori}(c) we plot the trade pattern $N^{j}(t)$ of  Merrill Lynch on Tuesday 08/21/01, and
also in Fig. \ref{MER4day} for the following three days, demonstrating that the Omori relaxation can persists
for several days. 

 The abrupt change in the curvature of $N(t)$ illustrates  the volatility clustering which begins around the time of the announcement $T$, corresponding to the vertical line at $t=285$ minutes
in Figs.  \ref{Omori}(a-c). 
For comparison, we find that the average $\langle N(t) \rangle$ time series calculated from all days without FOMC meetings is approximately linear with time throughout the entire day, 
indicating that the sudden increase in excess volatility before and after announcement  times $T$ results from the FOMC news. 
Volatility clustering in financial data sampled at the 1-minute scale persists for
several months, with a significant crossover in the observed power-law autocorrelations occurring around $600$ minutes
($\approx 1.5$ days) \cite{timescales, markettimeseries1, markettimeseries2}. 

In order to compare the dynamics before and after the announcement, we first separate the intraday time
series $N(t)$ into two time series $N_{b}(t \vert t < T)$, and $N_{a}(t \vert t > T)$. Then,  to treat the dynamics
symmetrically around the $i^{th}$ intraday announcement time $T_{i}$ \cite{SornettePNAS, ShortTermRxn}, we define the
displaced time $\tau = \vert t- T_{i}
\vert \geq 1$ as the temporal distance from the minute $T_{i}$ \cite{powlawfit}. As an illustration,  we plot $N(\tau)$ in  Fig.
\ref{Ntau}  for the 4 corresponding $N(t)$ curves exhibited in Fig. \ref{Omori}(a). We
then employ a linear fit  to both $N_{b,i}(\tau)=N_{i}(T_{i})- N_{i}(\vert t- T_{i}\vert)$ and $N_{a,i}(\tau)
=N_{i}(t-T_{i})- N_{i}(T_{i})$ on a log-log scale to determine the Omori power-law exponents $\Omega_{b}$ before the
news and $\Omega_{a}$  after the news.
 In analogy, we define the amplitude $\beta$ before as $\beta_{b}$ and after as $\beta_{a}$, as defined in
Eq.~(\ref{equation:Ncum}).

Typically $\Omega_{a}>0$, which reflects the pronounced increase in the rate of events above the volatility threshold
$q$ after the time of the announcement. 
We also
observe $\Omega < 0$, which corresponds to a time series in which the pre-shocks or after-shocks farther away  from the
announcement   (for large $\tau$) are dominant over the volatility cascade around time $\tau \approx 0$. For comparison, $n(\tau)$ is constant for stochastic
processes with no memory, corresponding to $\Omega \equiv  0$. Hence, for an empirical value $\Omega \approx 0$, the  rate $n(\tau)$
 is indistinguishable from an exponential decay  for $\tau < \overline t$, where $\overline t$ is the
characteristic exponential time scale. 
However, for larger values of $\Omega$, the exponential and power-law response curves are distinguishable, especially
if several orders of magnitude in $\tau$ is available.

 For all meetings analyzed, we find that $\Omega \equiv \Omega(q)$ increases with $q$, meaning that the relatively large aftershocks decay more quickly than
the relatively small aftershocks. Hence, the largest volatility values cluster around the announcement  time $T$. For
comparison, 
$\Omega(q)$ values are calculated in  \cite{OmoriLillo} using $q=4,5,6,7$ and in \cite{OmoriWeber} using $q=3,4$ for large financial crashes. 
For our data set, the cumulative probability $P(v> q)$ that a given volatility value is greater than volatility threshold $q$ is  $P(v > 3) = 0.18$ and $P(v > 5) = 0.087$. 
Furthermore, we reject the null hypothesis that $q>5$ volatilities are distributed evenly across all days, finding that $5\%$ of the
volatility values greater than $q=5$ are found on FOMC meeting days,
whereas only $4\%$ are expected under the null hypothesis that  large volatilities are distributed uniformly across all trading days. 
The $25\%$ increase for $q=5$ indicates
 that FOMC meetings days are more  volatile than other days at the $\alpha \approx 0$ significance level.
We also observe that the amplitudes of the Omori law generally obey the inequality $\beta_{b} <
\beta_{a}$, resulting from the large response immediately following the news. 

Although we focus mainly on price volatility $v(t)$ in this paper, we also observe Omori dynamics in the high-frequency
volume time series $\omega(t)$, defined as the cumulative number of shares traded in minute $t$. In Figs.~\ref{Alpha}
(a-d) for the S\&P 100 and Figs.~\ref{Alpha} (e-h) for the bank sector, we compare the average of Omori exponents $\Omega_{b}$ and $\Omega_{a}$
for both volatility and volume dynamics, and for volatility threshold value $q=3$. We compute the average Omori exponents using  two
averaging methods, the ``individual" method and the ``portfolio" method.

To analyze the time series $N_{a,i}$ after the announcement $i$, we first average the  exponents $\Omega^{j}_{a}$
obtained for each individual stock  $j$, yielding $\langle \Omega_{a} \rangle$. This ``individual" method provides an error bar
corresponding to the sample standard deviation $\sigma (\Omega_{a})$. The second ``portfolio" method determines  a
single $\Omega_{a}$ from $N(t)$ in Eq.~(\ref{equation:AveNcum}).
Comparing the open-box (individual method) and closed-box (portfolio method) symbols in  Fig. \ref{Alpha}, we observe
that both methods
 yield approximately the same average value of $\Omega_{a}$. Note that for the subset  $i$ =
\{1,4,8\} of the unscheduled FOMC meetings,  $\Omega_{a}$ is smaller than usual, capturing the intense
activity following surprise announcements.  Hence, unexpected FOMC announcements can produce an inverse
Omori law exhibiting convex  relaxation ($\Omega_{a} < 0$) over a short horizon if the news contains a large amount of
inherent surprise.
 The 8th meeting corresponds to the opening of the markets after Sept. 11, 2001.

For the time series $N_{b,i}$ before the announcement $i$, individual stocks often do not have sufficient activity to
provide accurate power-law fits. Hence, to estimate the sample standard deviation $\sigma (\Omega_{b})$, we produce
partial combinations, $\langle N(\tau) \rangle_{b,i} \equiv \frac{1}{M}\sum_{j=1}^{M} N^{j}_{b,i}(\tau)$ using $M \equiv 5$.
 We then compute a standard deviation $\sigma (\Omega_{b})$ from the  $\Omega_{b}$ values calculated from $\langle N(\tau) \rangle_{b,i}$. 
The  $\sigma (\Omega_{b})$ values correspond to the error bars for $\langle \Omega_{b} \rangle$ in
Fig.~\ref{Alpha}. 

We also compute  a single $\Omega_{b}$ value from the portfolio average $N_{b,i}(\tau)$, which corresponds to  the limit $M = S$. The
 values of   $\Omega_{b}$ using the two methods are consistent. 
Interestingly, the values of $\Omega_{b}$
calculated from volume data are all close to zero. However,  using the Student T-test we reject the null hypothesis 
that each average value $\langle \Omega_b \rangle$ is equal to zero at the $\alpha = 0.01$ significance level for 15 out of 17 dates.

 Fig. \ref{Alpha}  shows the range of  $\Omega$ values for each of the 19 FOMC meetings we analyze. There are 8 panels comparing the  $\Omega$ values  (i) between
the dynamics before and after $T$, (ii) between the  volatility and volume dynamics,  and (iii) between set of all stocks comprising the $S\&P100$ and the set of stocks  comprising the banking sector. 
We hypothesize that the differences in the Omori $\Omega$ values,  before and after the announcement, are related to the anticipation and 
perceived surprise of the FOMC news. Furthermore, for the dynamics after the news, we find anomalous negative $\Omega_{a}$ values  for two surprise FOMC announcements  $i=1$ and $i=8$.
Also, we find that volume $\Omega$ values  are more regular across all meeting events, suggesting that volume and price volatility contain distinct market information \cite{Earnings, VolPriceRxn}.

In order to find potential variations in the response dynamics for different stock sectors, 
 in Fig.~\ref{SectorOmega}(a) we compare the  $\Omega_{a}$ values after the announcement
  for $5$ approximately equal-sized sectors using volatility threshold $q=3$. We observe that the differences in the average values of the
sectors are fairly small, indicating a broad market response. We also observe that the technology sector
(Tech.), composed of hardware, software, and IT companies, often has the largest average $\Omega_{a}$ value. Larger exponents, which correspond to shorter relaxation times, could result from the intense trading in the
Tech sector during the Tech/IT bubble, which peaked in the year 2000. 
In follow-up analysis, we find in \cite{marketshocks} that stocks with higher trading activity, quantified as the average number of transactions per minute, have larger $\Omega^{j}$ 
in response to market shocks, and thus, faster price discovery.
In order to  compare the variation in the
individual values of $\Omega^{j}_{a}$, we  plot the pdf of exponents for all  stocks and meetings in
Fig.~\ref{SectorOmega}(b) using the  shifted variable $x^{j} \equiv \Omega^{j}_{a,i} -  \Omega_{a,i}$. We conclude from a Z-test at the $\alpha = 0.0005$ significance level
that Tech sector Omori exponents are larger on average, $\langle x \rangle_{Tech} > \langle x \rangle_{SP100}$.

Motivated by the metric $\Theta_{i}$ defined in Eq.~(\ref{thetadef}), which quantifies speculation and anticipation in
the market preceding FOMC meetings, we now develop 
a second metric to describe surprise through the change in market speculation following the announcement. 
This metric $\Delta_{i}$ compares the anticipation leading up to the announcement with the revised speculation following
the FOMC decision. This can be quantified through the relative change in $\delta(t)$, which provides a rough measure of
the market stress that is released in the  financial shock. Qualitatively, $\Delta_{i}$ relates  the 
average value of the spread before and after the $i^{th}$ scheduled  meeting. We define,
\begin{equation}
\Delta_{i} \equiv  \Big( \overline{\delta(t)}_{i, a}-\overline{\delta(t)}_{i, b} \Big) \times S(\Delta R_{i})
\end{equation}
\begin{equation}
\equiv \Big( \frac{\sum \delta(t_{i}-\Delta t) \ w(\Delta t)}{\sum w(\Delta t)} - \frac{\sum \delta (t_{i}+\Delta t)\
w(\Delta t)}{\sum w(\Delta t)}\Big) \times S(\Delta R_{i}) \ ,
\label{abstheta}
\end{equation}
where the sum is computed over the range $\Delta t \in [1, L_{2}]$ trading days, with $L_{2}=15$ trading days and
$\lambda_{2} = 10$ trading days. 
The factor $ S(\Delta R_{i})=1$ when the Fed increases or maintains the Target rate $R(t)$, while $S(\Delta R_{i})=-1$
when the Fed decreases the Target rate. 

 In Figs. \ref{ThetaIntraday} (a-d) we relate the amplitudes $\langle \beta_{b} \rangle$ and $\langle \beta_{a} \rangle$, and also
 the exponents $\langle \Omega_{b} \rangle$ and $\langle \Omega_{b} \rangle$ to the speculation metric $\Theta$ and the surprise metric $\Delta$. 
  We observe that larger $\Theta$ and larger $\Delta$ are related to larger amplitude
$\langle \beta_{b} \rangle$ quantifying the preshock dynamics. 
However, we do not find a statistically significant relation between $\Theta$ or $\Delta$ and the aftershock parameters, 
suggesting that the relaxation dynamics following FOMC news are less predictable.
Nevertheless, the aftershock dynamics are consistently more pronounced, with $\langle \beta_{a} \rangle > \langle \beta_{b} \rangle$.
We interpret  Figs. \ref{ThetaIntraday}(a) and (c) as follows: when $\Theta<0$, corresponding to ``good" market sentiment and possible rate increase,
the dynamics before the announcement have small $\beta_{b}$ and small $\Omega_{b}$ reflecting low activity. After the announcement, the values of
 $\beta_{a}$ and $\Omega_{a}$ increase, corresponding to a  fast response of medium size. In the case of $\Theta>0$, corresponding to ``bad" market sentiment resulting from speculation of a rate cut,
the dynamics before the announcement have large $\beta_{b}$ and large $\Omega_{b}$, corresponding to a strong but quick buildup of volatility. After the announcement, 
the dynamics have large $\beta_{a}$ and small $\Omega_{a}$, corresponding to a strong and lasting relaxation dynamics. 
The interpretation of Figs. \ref{ThetaIntraday}(b) and (d) is similar to the interpretation of Figs. \ref{ThetaIntraday}(a) and (c), in that both 
surprise ($\Delta >0$) and expected ($\Delta \approx 0$) bad news,   correspond to a stronger and longer-lasting relaxation dynamics.

\section{Discussion} Information flows through various technological avenues, keeping the ever-changing world
up-to-date. All news carries some degree of surprise, where the perceived magnitude of the news certainly depends on the
recipient. 
In financial markets, where speculation on investment returns results annually in billions of dollars in transactions,
news plays a significant role in perturbing the complex financial system both on large and small scales, reminiscent of
critical behavior with divergent correlation lengths \cite{3pillars}. 
Perturbations to the financial system are easily transmitted throughout the market by the long-range interactions that
are found in the networks of market correlations \cite{marketcorr0, marketcorr1,marketcorr2}. Afterwards, the effects of
the perturbation may persist via the long-term memory observed in volatility time series
\cite{ timescales, markettimeseries1, markettimeseries2}, with fluctuation scaling obeying the emperical Taylor's law 
\cite{taylorslawstocks, scalingtempcorr}. 

We have shown that the Omori law describes the dissipation of information following the arrival of Federal Open Market
Committee (FOMC) news. This type of relaxation is consistent with the substructure of financial crash aftershocks observed on various scales \cite{OmoriWeber}. In particular, we systematically study the dynamical response of the stock market to
perturbative information in the form of a Federal Reserve FOMC interest rate announcements, which can be expected
(scheduled) or unexpected (as in cases of emergency). 

 In the case of unexpected news, as in Fig. \ref{Omori}(d), a pronounced response may result from reduced 
market liquidity, since traders do not have ample time to prepare and adjust \cite{MacroNews2}.  Our findings suggest that the dynamics of ``rallies" based on
other forms of news, such as earning reports, upgrades and downgrades of stocks by major financial firms, unemployment
reports, merging announcements etc., might also be governed by the Omori law with parameters that depend on the type of news. The impact of 
macroeconomic news has been analyzed for foreign exchange markets \cite{MacroNews2}, where it is found that high levels of volatility  are present following 
both scheduled and surprise news. 

 According to the efficient market hypothesis \cite{noEMH}, the time scale over which news is incorporated into 
prices  should be very small. However, consistent with previous studies, we find ``market underreaction" \cite{sentiment}
evident in the  finite time scale (found here to be at least 1 trading day) over which the volatility aftershocks are significant. 
Moreover, we quantify the dynamics before and after, and show that the Omori parameters are related to investor sentiment \cite{sentiment}, here measured
by comparing the 6-month the Treasury Bill and the Federal Funds rates. 
   
It is also conceivable that  Omori law decay of market aftershocks 
also exists in  the  traded volume time series and  the bid-ask spread time series \cite{ShortTermRxn,
limitorderOmori} . Recently, Joulin {\it et al.} 
\cite{information} use a similar method to describe the relaxation of trading following news streaming from feeds such
as Dow Jones and Reuters, and compare to the relaxation following anomalous volatility jumps. 
Joulin {\it et al.} \cite{information}  also find Omori law relaxation, with exponent $\Omega_{a} \approx 1$  following a
news source, and $\Omega_{a} \approx 0.5$ following an endogenous jump; interestingly, they find that the amplitude of the
Omori law is larger for news sources than for endogenous jumps. For further comparison, Weber {\it et al.}
\cite{OmoriWeber} find $\Omega_{a} \approx 0.69$ for the 38 days following the market crash on September 11, 1986. One
distinct difference between these studies, is the source of the news: Joulin {\it et al.} pool together thousands of 
news sources, some possibly pertaining to only a single stock; we focus on one particular type of news, the FOMC Target rate
decision, which has a broad  impact on the whole market and economy. It is possible that the difference between
{\it anticipated} news and {\it idiosyncratic} news is the important criterion to consider when analyzing market
response functions in relation to exogenous events. Here, we find novel dynamics before anticipated announcements.

 In the case of FOMC news, speculation can be quantified by measuring the relative difference between
 the effective Federal Funds rate and the Treasury Bill in the weeks leading up to a scheduled meeting. We develop a
speculation metric, $\Theta$, and relate it to $V$, the volatility on the day of the meetings, finding that the market
behaves more erratically when the Treasury Bill predicts a decrease in the Federal Funds Target rate.
A rate decrease often occurs in response to economic shocks, whereas a rate increase is often used to fight inflation.
Hence, the asymmetric response in Fig. \ref{ThetaDaily} to rising and falling rates is consistent with the ``sign effect", where 
it has been found   that bad news causes a larger market reaction than good news \cite{MacroNews2}, and that the asymmetry may result from
the increased uncertainty in expectations among traders. 

We analyze the four Omori-law parameters $\Omega_{b}$, $\Omega_{a}$, $\beta_{b}$ and $\beta_{a}$ calculated for 19 FOMC meetings.
  We conjecture that the Omori-law parameters are related to the
 market's speculation, anticipation and surprise on the day of the FOMC meeting. 
 In order to quantify speculation of rate cuts and rate increases, we define the measure $\Theta$, which is the  relative spread
between the Treasury Bill and the Federal Funds rates, before the meeting. 
In order to quantify surprise, we develop $\Delta$, which measures the change in the  relative spread
between the Treasury Bill and the Federal Funds rates, before and after the meeting. We relate both $\Theta$ and $\Delta$ to the dynamical
response of the market on the day of the meeting. We find that relatively small $\Omega$ values and relatively large amplitude $\beta$
values, corresponding to longer relaxation time and large response, follow from ``bad" news, as in the case of the market reaction to the  World Trade Center attacks in 2001. 
 In all, these results show that markets relax according to the Omori law following large crashes
and Federal interest rate changes, suggesting that the perturbative response of markets belongs to a
universal class of Omori laws, independent of the magnitude of news.

\section{Acknowledgements} We thank L. DeArcangelis and M. Levy for helpful suggestions and NSF for financial support.

\begin{widetext} 

\begin{figure}[b]
\centering{\includegraphics[width=0.7\textwidth]{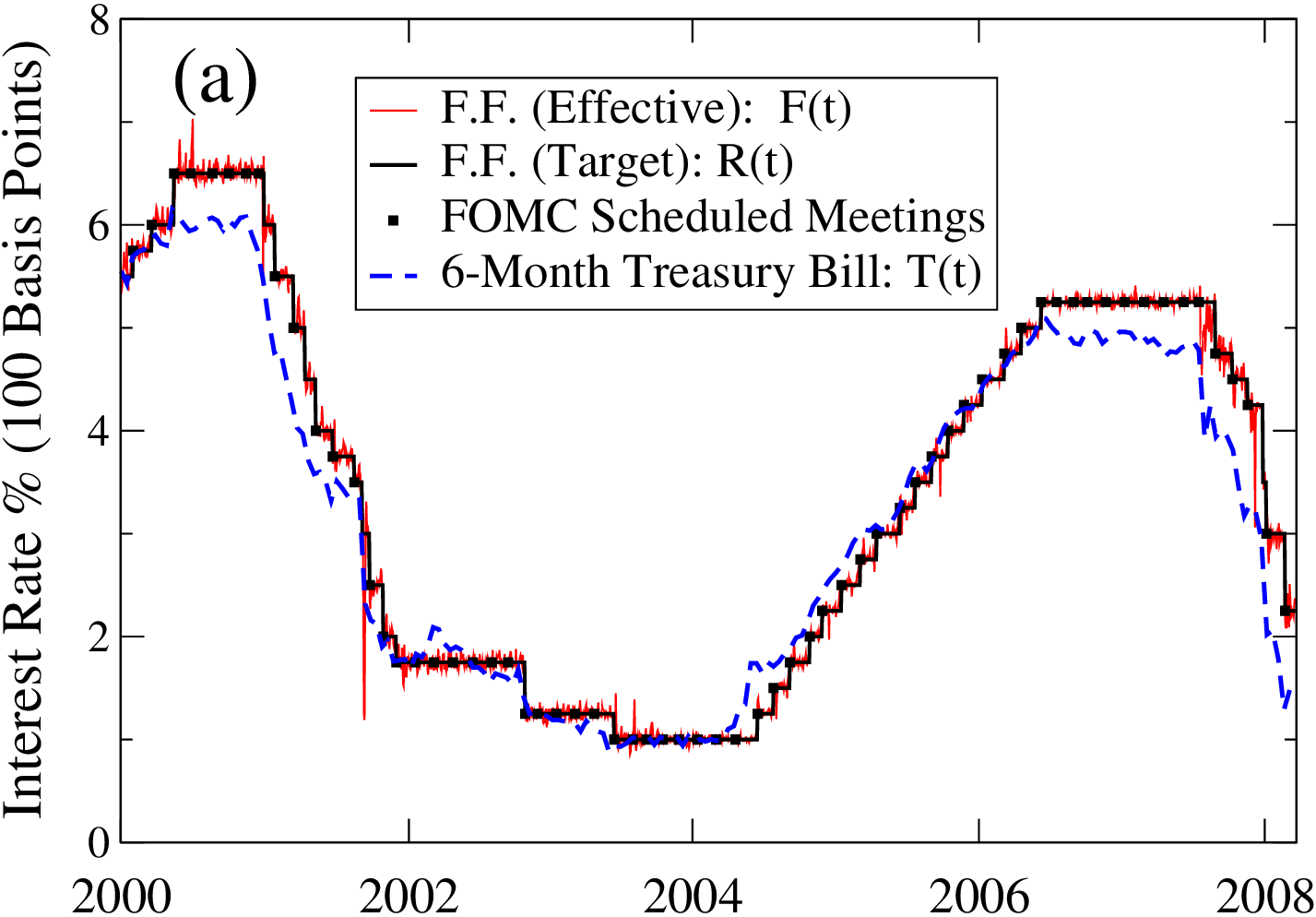}}\\

\centering{\includegraphics[width=0.7\textwidth]{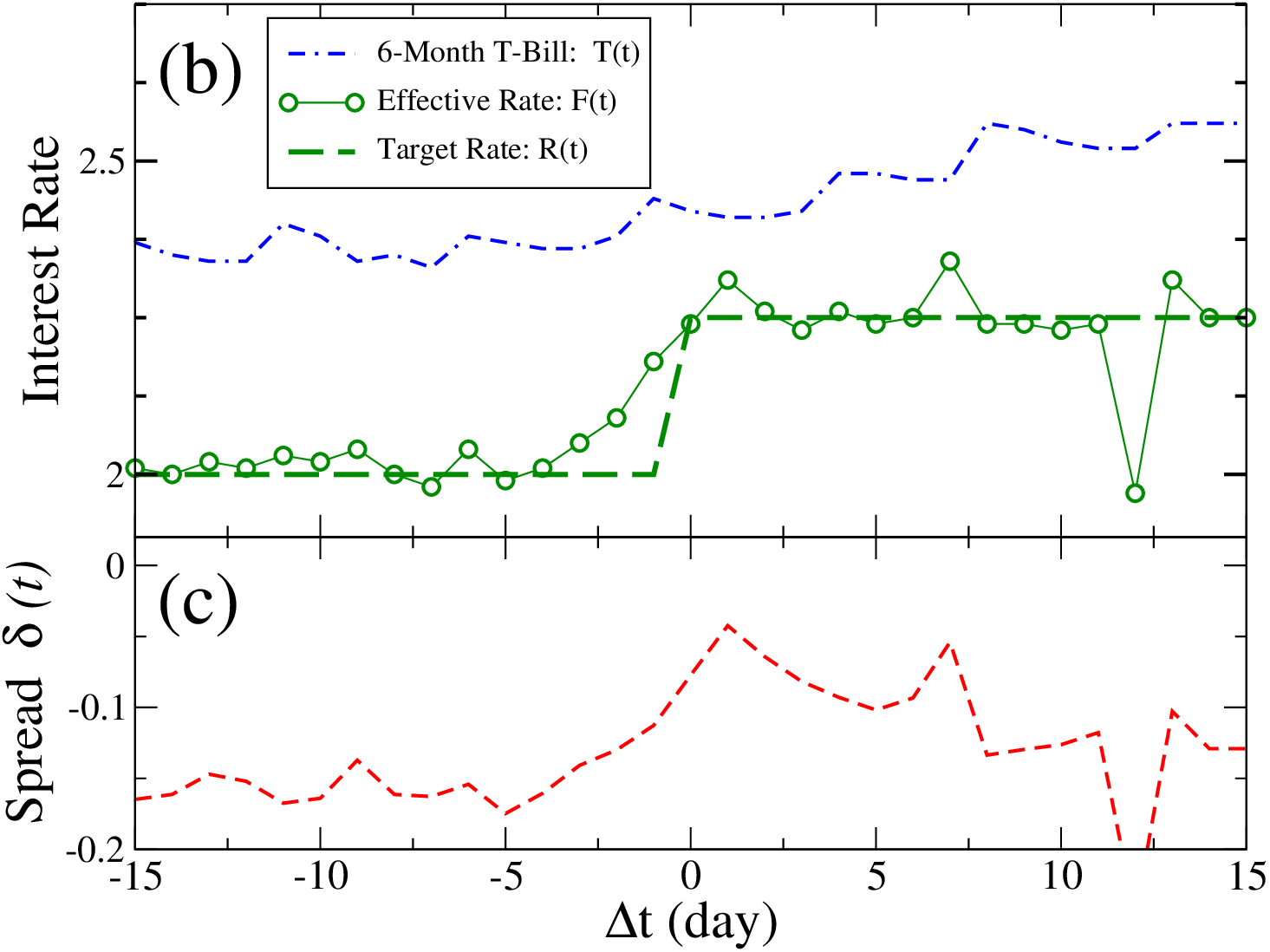}}
\caption{ \label{deltaexample}  (Color online) An illustration of the close relation between the Treasury Bill and the
Federal Funds rate. (a)  Time series of the Federal Reserve Target rate, $R(t)$, and the Federal Reserve Effective rate,
$F(t)$, for Federal Funds (F.F.) dating from Jan. 2000 to Apr. 2008. The 6-Month Treasury Bill, $T(t)$, closely follows
the effective rate, with speculation about future changes causing deviations in the relative values. United States
Treasury Bills carry little risk, and are considered to be one of the most secure investments. (b) A typical
illustration of the Federal Funds Effective rate and the Treasury Bill, where both gravitate around the Federal Funds
Target rate. The change in the relative spread $\delta(t)$, defined in Eq.~(\ref{deltat}), between the Treasury bill and
the Federal Funds Effective rate, indicates changes in market speculation. (c) The relative spread, $\delta(t)$, 15 days
before and 15 days after the scheduled FOMC meeting on Dec. 14, 2004, which corresponds to  $\Delta t=0.$ Note that the
average value of the relative spread increases after the announcement, indicating a shift in market consensus and
speculation.
 }
\end{figure} 

\begin{figure}
\centering{\includegraphics[width=0.7\textwidth]{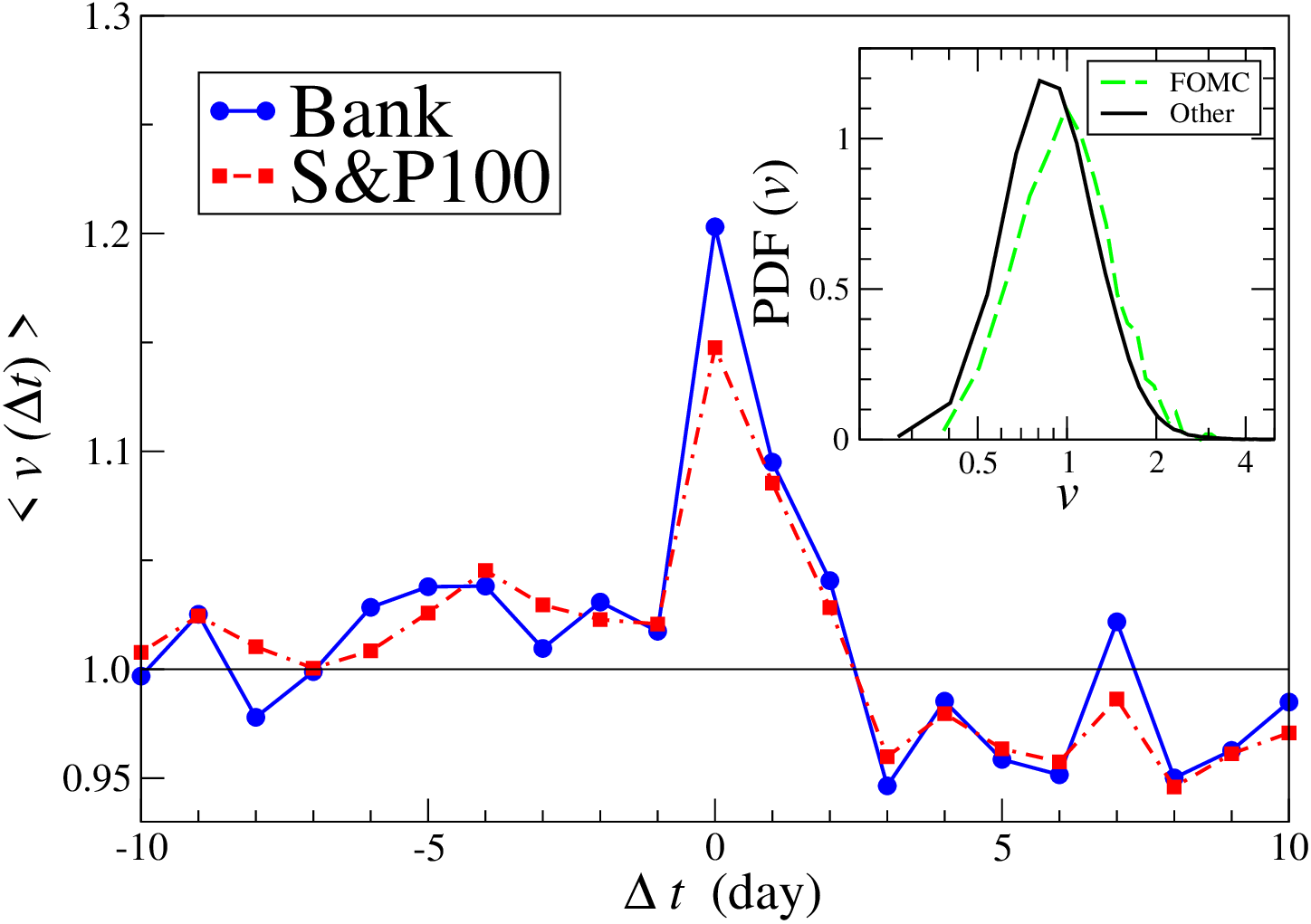}}
 \caption{ \label{Daily}  (Color online) Average daily volatility trend defined in Eq.~(\ref{eq1}) exhibits increased
market volatility on the day of FOMC meetings, corresponding to $\Delta t=0$.
 ``Bank" refers to the portfolio of 18 stocks that belong to the S\&P100. There is a 15-20\% increase in 
volatility on days corresponding to FOMC meetings. Standard deviation $\sigma(v(\Delta t))\approx 0.4$ can be assigned to
each data point in the time series, and
are calculated by randomizing the daily volatility time series of each company. (Inset)  Probability density
function (pdf) of normalized volatility
$v \equiv r(t) / \langle r \rangle$, where the quantity $r(t) \equiv \ln  (p_{hi}(t)/p_{low}(t))$ is the range of the
price time series of a given stock for a particular day. We plot the pdf of volatility values for the 
S\&P100 on the set of days with FOMC meetings and for the set of all ``other" days. The distributions are approximately
log-normal, with a shift towards higher average volatility on FOMC days. The average values for the two data sets are 
$\langle v \rangle_{FOMC} = 1.12$ and $\langle v \rangle_{other} = 1.00$.
 }
\end{figure}

\begin{figure}
\centering{\includegraphics[width=0.7\textwidth]{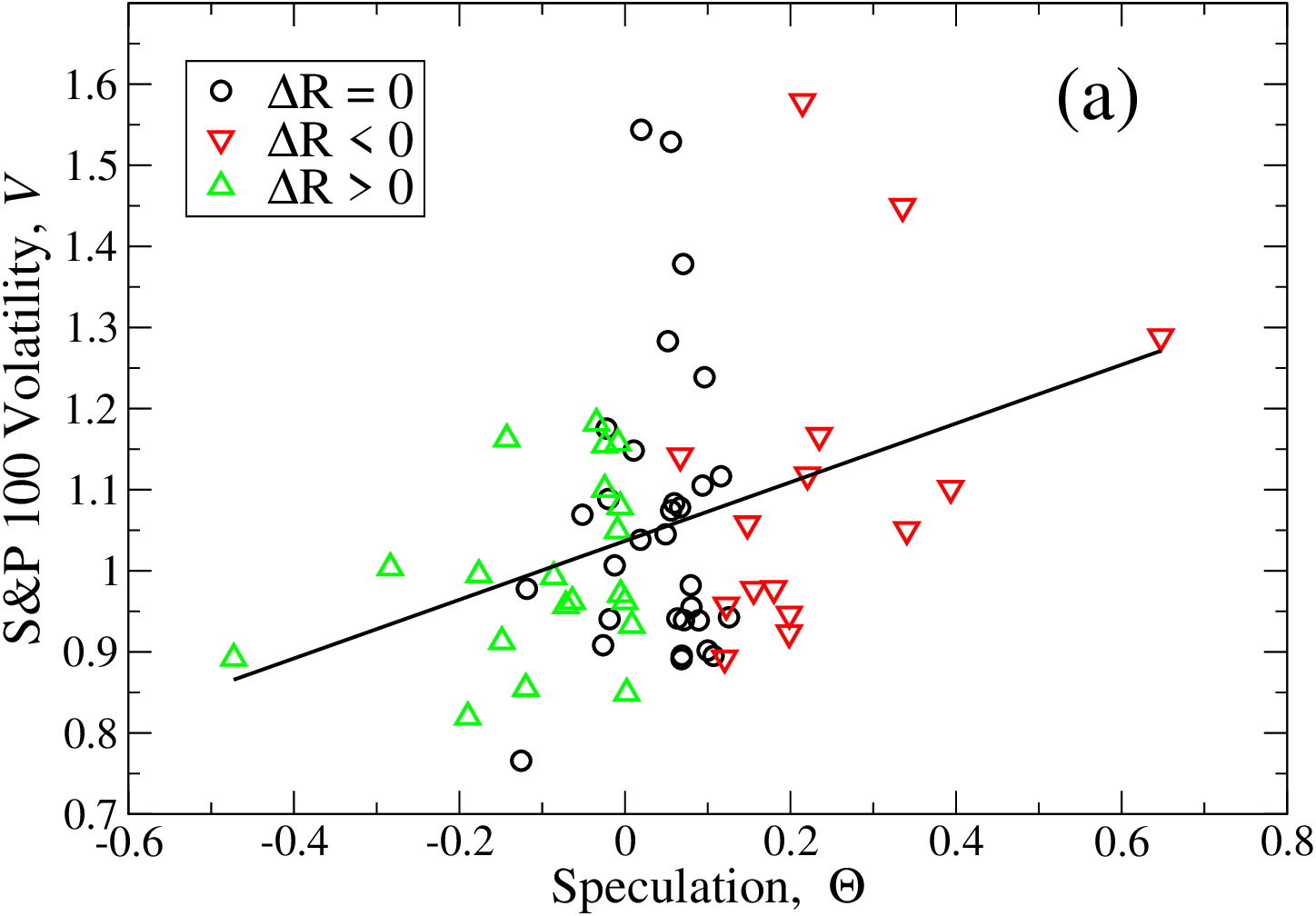}}\\

\centering{\includegraphics[width=0.7\textwidth]{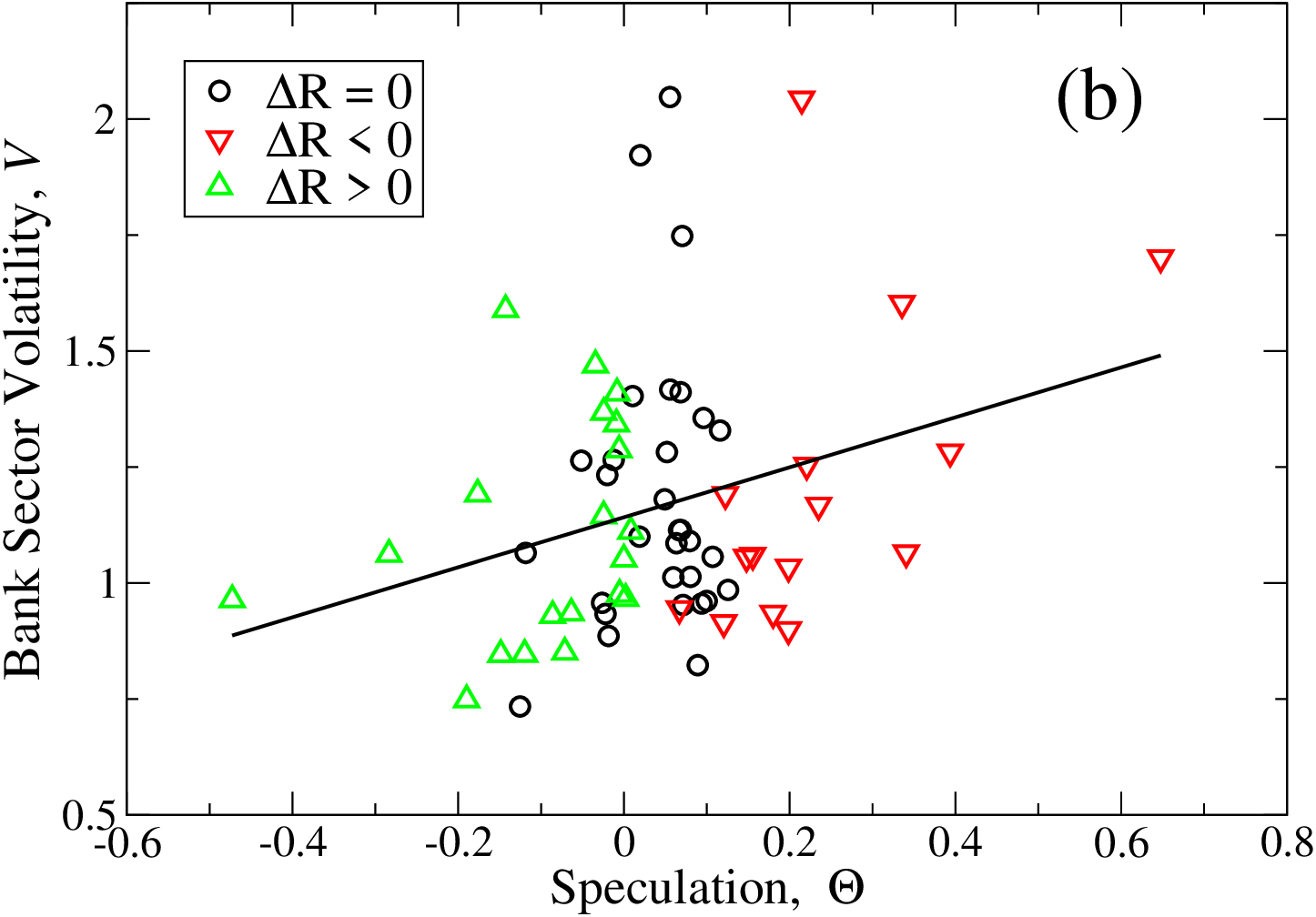}}
 \caption{ \label{ThetaDaily} (Color online) Demonstration of the relation between speculation of interest rate change
and
 market volatility in the S\&P100 and for the subset of {\it Banking} stocks. We relate 
$\Theta_{i}$,
the speculation in the market before a FOMC meeting defined in Eq.~(\ref{thetadef}),  to the market volatility $V_{i}$
defined in Eq.~(\ref{V0}). 
A large absolute value of $\Theta_{i}$ reflects the high probability that an interest rate change will happen. 
Interestingly, there are many instances where $\Theta_{i} \approx 0$, followed by  large volatility. These values
correspond to FOMC decisions to maintain interest rate levels ($\Delta R =0$), and suggest a fundamental difference in
the dynamics following decisions to change versus decisions not change the Federal Funds Target rate.
Also, there is an underlying symmetry in $\Delta R$, since  in the  case of  either a rate increase
or
a rate decrease, the FOMC also  has the option of no increase. Hence, $\Delta R =0$ can occur as either ``good" or
``bad" news, whereas typically decisions of $\Delta R>0$ reflect situations with  positive market sentiment whereas
decisions of $\Delta R<0$ reflect situations with negative market sentiment. Hence, the asymmetry in market volatility
is consistent with the ``sign effect" \cite{MacroNews2}.
 Although the
correlation between $\Theta_{i}$ and $V_{i}$ is dominated by residual error, it is 
nevertheless supporting that the regression captures the crossover at $(\theta, V) = (0,1)$. Including all data points, 
the regression correlation coefficient is $r^{2} = 0.34$,  and the slope of the regression is $m = 0.36\pm 0.13$ for panel
(a) and  $r^{2}=0.30$  and  $m =0.54\pm 0.22$ for panel (b). Restricting data points corresponding only to interest rate
changes (red and green triangles), 
$r^{2} = 0.48$ and $m= 0.37\pm 0.12$ for panel (a) and $r^{2}=0.40$ and $m =0.53\pm 0.21$ for panel (b) (this second
regression is not shown and is indistinguishable from the regression including all data points). All linear regressions
pass the ANOVA (analysis of variance) F-test, rejecting the null hypothesis that $m=0$ at the $\alpha = 0.05$ significance level.
 }
\end{figure}

\begin{figure}[t]
\centering{\includegraphics[width=0.7\textwidth]{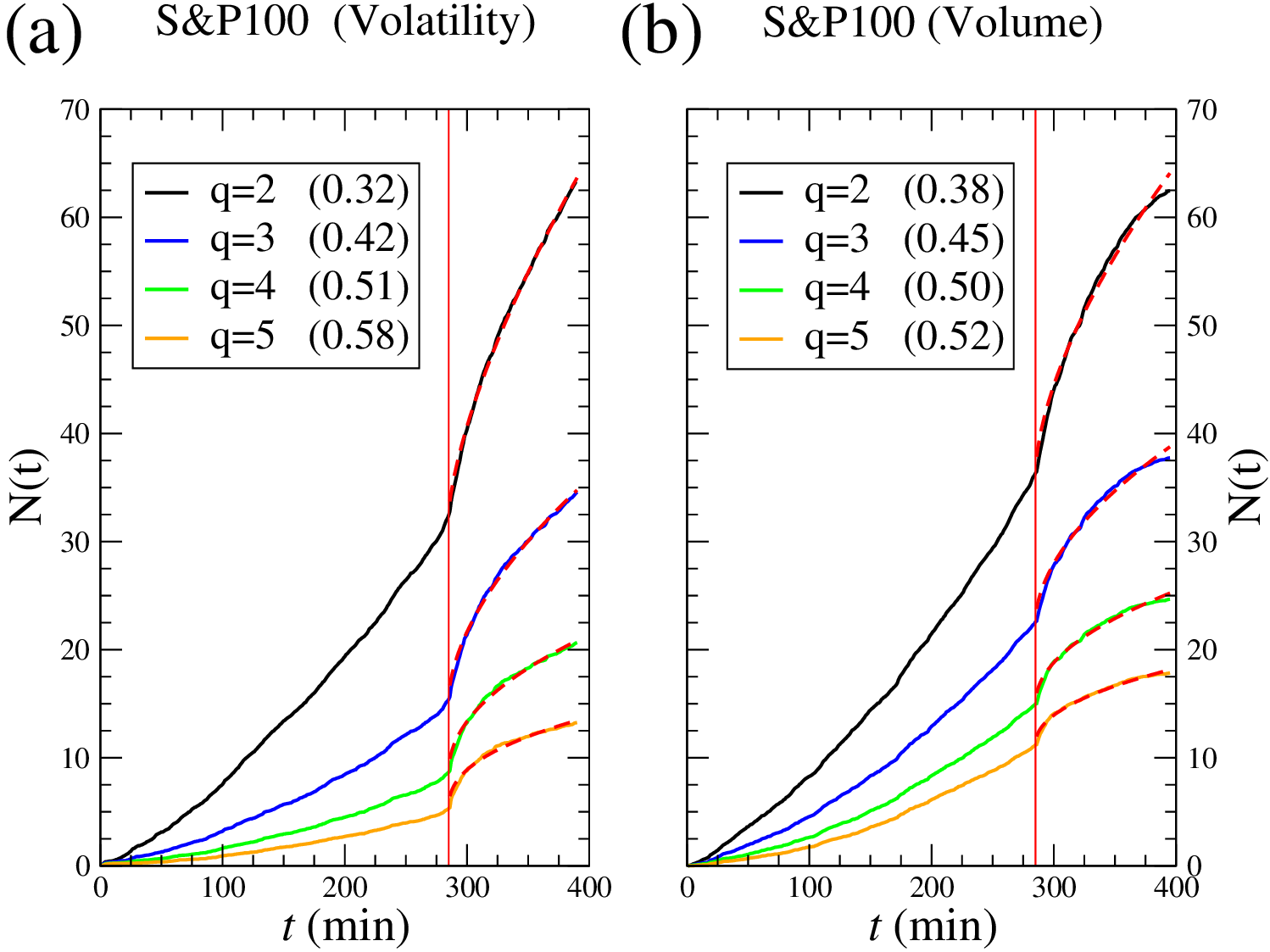}} \\
\bigskip

\centering{\includegraphics[width=0.7\textwidth]{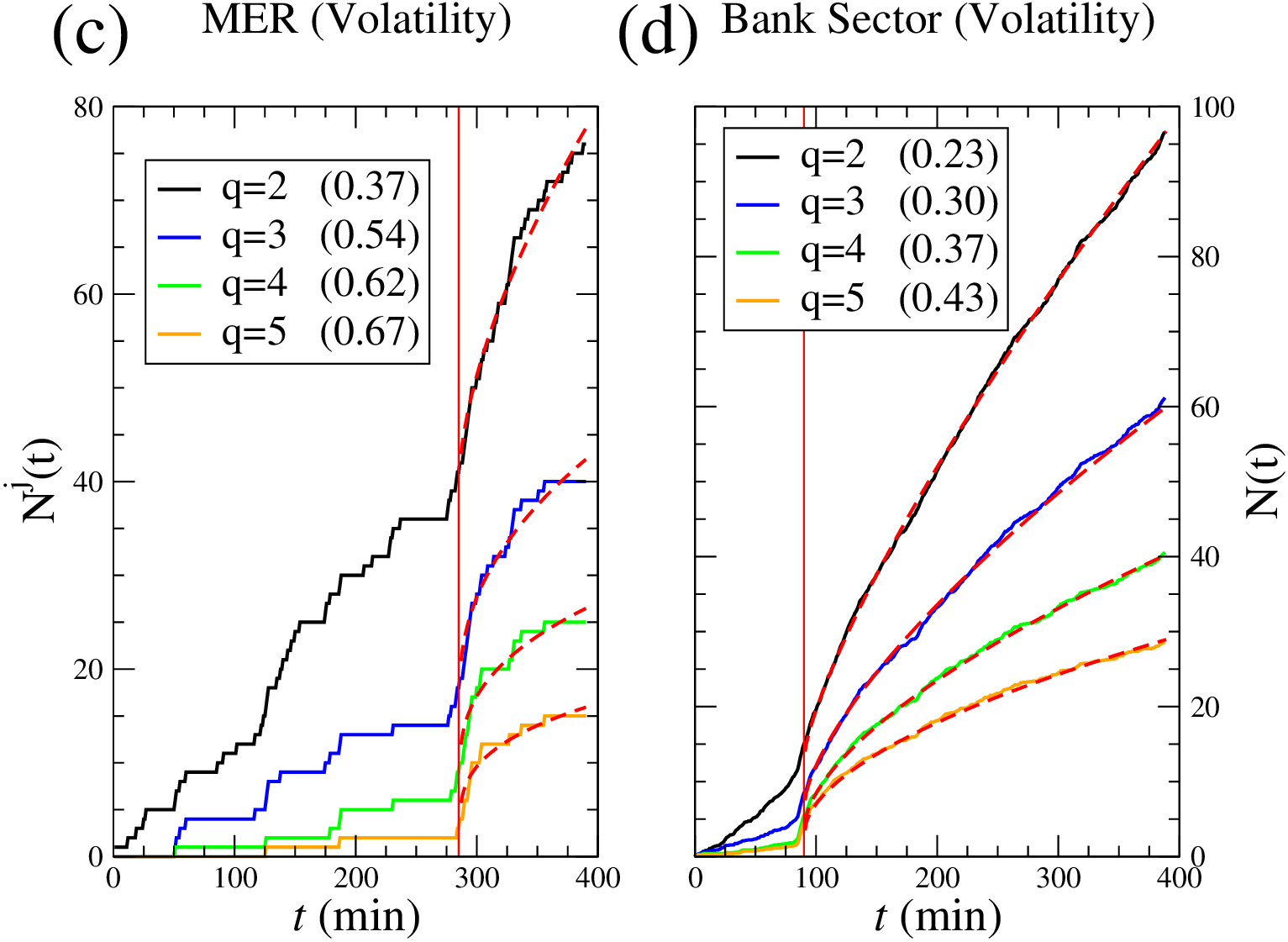}}
  \caption{ \label{Omori} (Color online) The cumulative volatility time series  $N(t)$  demonstrates 
Omori-law response dynamics,
  here  in response to FOMC announcements occurring at the time $T$ indicated by a vertical solid red line. 
  The abrupt change in the curvature of $N(t)$ around time $t \approx T$ illustrates  the increased volatility caused
by the announcement.
  The significant aftershocks which occur until the end of the trading day are consistent with  ``market underreaction"
\cite{MacroNews1,MacroNews2}.
   Market underreaction and other market inefficiencies can result from increased levels of uncertainty among traders
following market news \cite{sentiment}.
  Each time series $N(t)$ is calculated for a given volatility threshold $q$, where larger $q$ values 
  correspond to $N(t)$ curves with smaller amplitude (smaller rate of large volatility events). 
Panels (a,b,c) illustrate the dynamics around a scheduled announcement made at $T = 285$ minutes (2:15 PM ET).  For the
S\&P 100, we calculate $N(t)$ on  05/15/01  for (a) 1-minute volatility and (b) 1-minute total volume
 using Eq.~(\ref{equation:AveNcum}). (c) We calculate  $N^{j}(t)$  for Merrill Lynch (MER) on  08/21/01. 
(d) The Omori law also occurs for {\it unscheduled} FOMC announcements, as illustrated for the {\it Bank} sector $N(t)$ 
on 04/18/01, when the surprise rate change was announced  at $T = 90$ minutes (11:00 AM ET),
 resulting in raised levels of volatility throughout the entire trading day. For panels (a-d), the dashed red lines are
power-law fits beginning immediately after the announcement, with the corresponding exponents $\Omega_{a} (q)$ appearing
in parentheses within the legends. 
} 
\end{figure}

\begin{figure}
\centering{\includegraphics[width=0.7\textwidth]{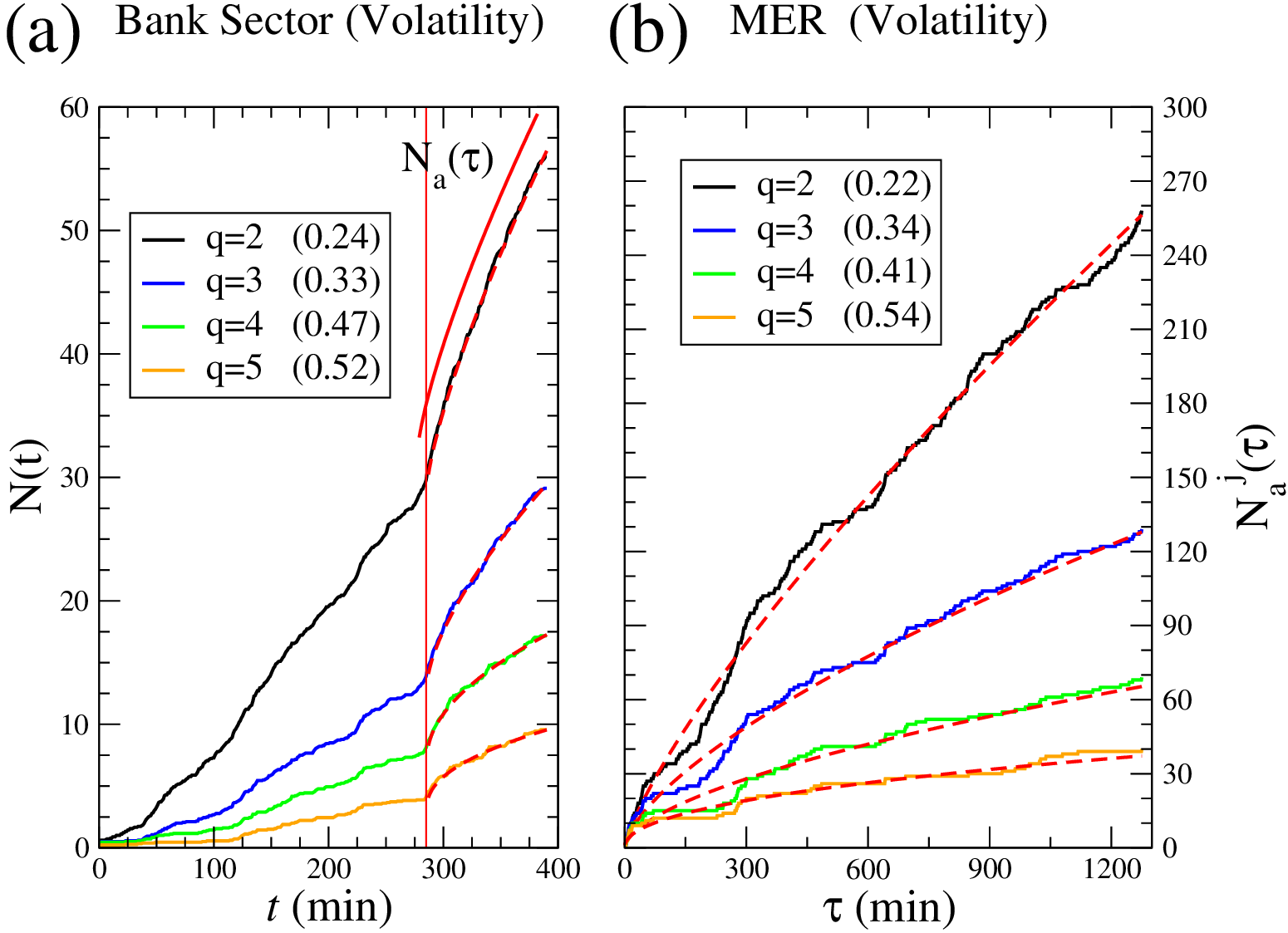}}
 \caption{
  \label{MER4day} (Color online)  The Omori law relaxation can extend for several days. We compare the Omori
exponents $\Omega_{a}(q)$ (indicated in legends) calculated for  (a) the time series $N_{a}(\tau)$ of the {\it Bank}
sector and (b) the time series  $N^{j}_{a}(\tau)$ of Merrill Lynch for the three days 
(1275 minutes) after the announcement on Tuesday 08/21/01 at 2:15 pm ($\tau = 0$ minutes corresponding to  $T = 285$
minutes).
For the remaining 3 days of the trading week, the Omori law relaxation corresponding to an individual stock (MER) is
quantitatively similar to the the Omori law relaxation of the {\it Bank} sector over the final  105 minutes of the
initial trading day. We do not use the {\it Bank} sector $N_{a}(\tau)$ over the same 1275-minute time period for
comparison because  ``opening effects" occurring during the first 60 minutes of each trading day 
makes power-law regression of conjoined $N_{a}(\tau)$ problematic. } 
\end{figure}

\begin{figure}[t]
\centering{\includegraphics[width=0.7\textwidth]{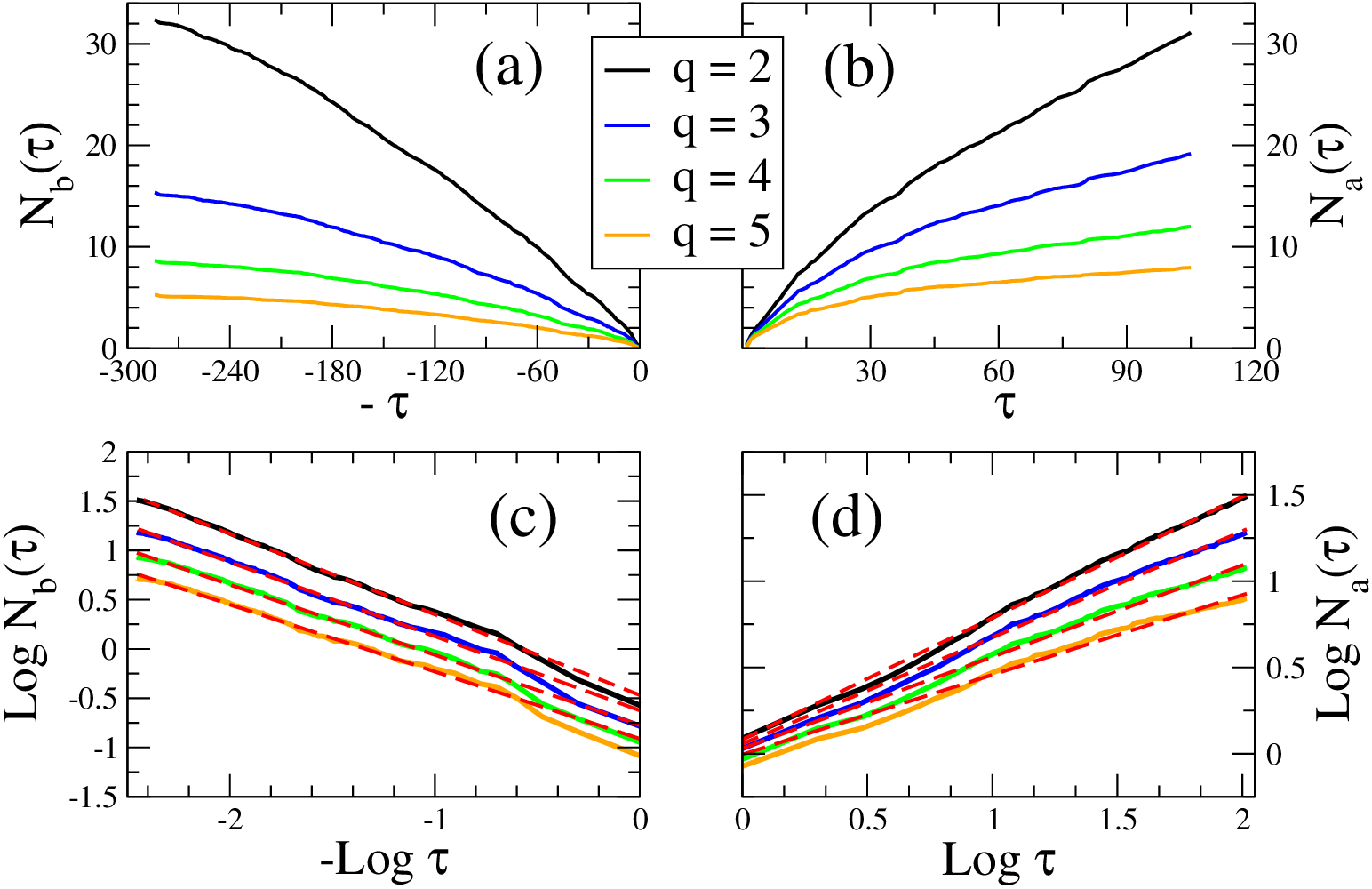}}
  \caption{ \label{Ntau} (Color online) An illustration of the method used to calculate (a) $N_{b,i}(\tau)=N_{i}(T_{i})-
N_{i}(\vert t- T_{i}\vert)$ and (b) $N_{a,i}(\tau) =N_{i}(t-T_{i})- N_{i}(T_{i})$   for each  intraday time series
$N_{i}(t)$. The  displaced time $\tau = \vert t- T_{i} \vert$ is defined  symmetrically around
  the time of the announcement  $T_{i}$. We plot the same data as in  Fig. \ref{Omori} ({\it a}), corresponding to the
announcement on 05/15/01 which occurred at  $T = 285$ minutes.
 Panels (c) and (d) show that 
$N_{b,i}(\tau)$ and $N_{a,i}(\tau)$ are approximately linear on logarithmic scale. } 
\end{figure}

\begin{figure}[t]
\centering{\includegraphics[width=0.7\textwidth]{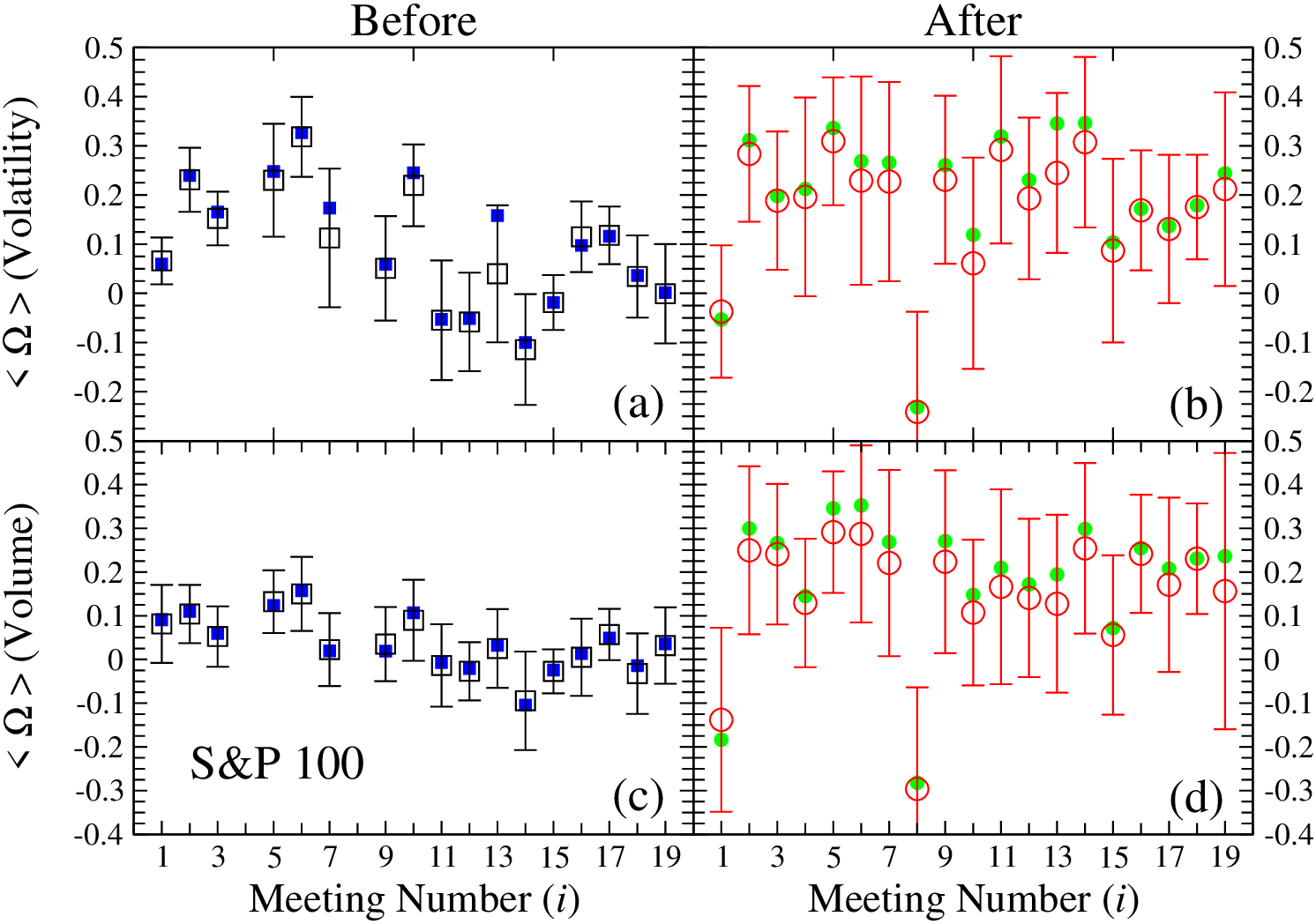}}\\

\centering{\includegraphics[width=0.7\textwidth]{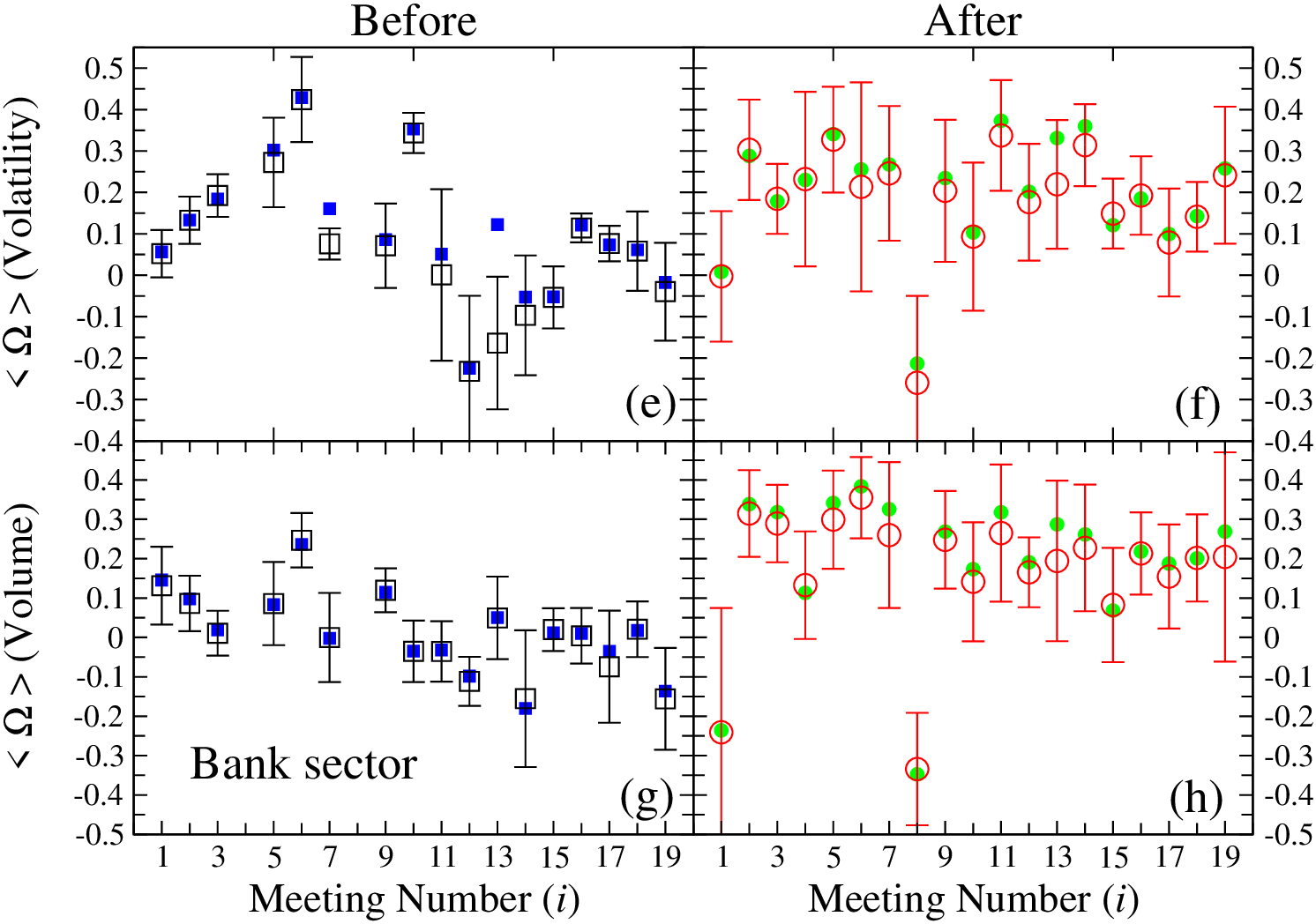}}
\caption{ \label{Alpha}  (Color online) Comparison of Omori law exponents for both volatility dynamics and  volume dynamics on the day of 19 FOMC meetings 
 during the two-year period Jan. 2001- Dec. 2002. Panels (a-d) correspond to the S\&P 100 and panels(e-h) correspond to
the bank sector. 
 The average value of $\Omega$ for the 16 scheduled FOMC meetings (excluding the 3 unannounced meetings  $i$ =
\{1,4,8\}) are: 
 (a) $ \overline{ \Omega}_{b}  = 0.10 \pm 0.13 $, (b) $ \overline{ \Omega}_{a}   =  0.24 \pm 0.08 $, (c) $ \overline{
\Omega}_{b}  = 0.04 \pm 0.07 $, (d) $ \overline{ \Omega}_{a}   = 0.24 \pm 0.07 $,  (e) $ \overline{ \Omega}_{b}   = 0.11
\pm 0.16 $, (f) $ \overline{ \Omega}_{a}   =  0.23 \pm 0.09 $, (g) $ \overline{ \Omega}_{b}   = 0.01 \pm 0.10 $, (h) $
\overline{ \Omega}_{a}   =  0.26 \pm 0.08$. The similarity in exponents for 1-minute volatility and 1-minute cumulative
volume suggest a universal underlying mechanism. Solid symbols ($\blacksquare$ and $\bullet$) refer to $\Omega$ computed
from $N(t)$.
 Open symbols ($\square$ and $\bigcirc$) refer to $\langle \Omega \rangle$ computed from $S$ individual Omori exponents
$\Omega^{j}$, with $S_{bank}=18$. Note the  relatively low value of $\Omega_{a}$ and $\langle \Omega_{a}\rangle$ for
unscheduled FOMC announcements $i=$ 1 and 8, which indicates that volatility rate following the
announcement increased throughout the day.} 
\end{figure}


\begin{figure}[t]
\centering{\includegraphics[width=0.7\textwidth]{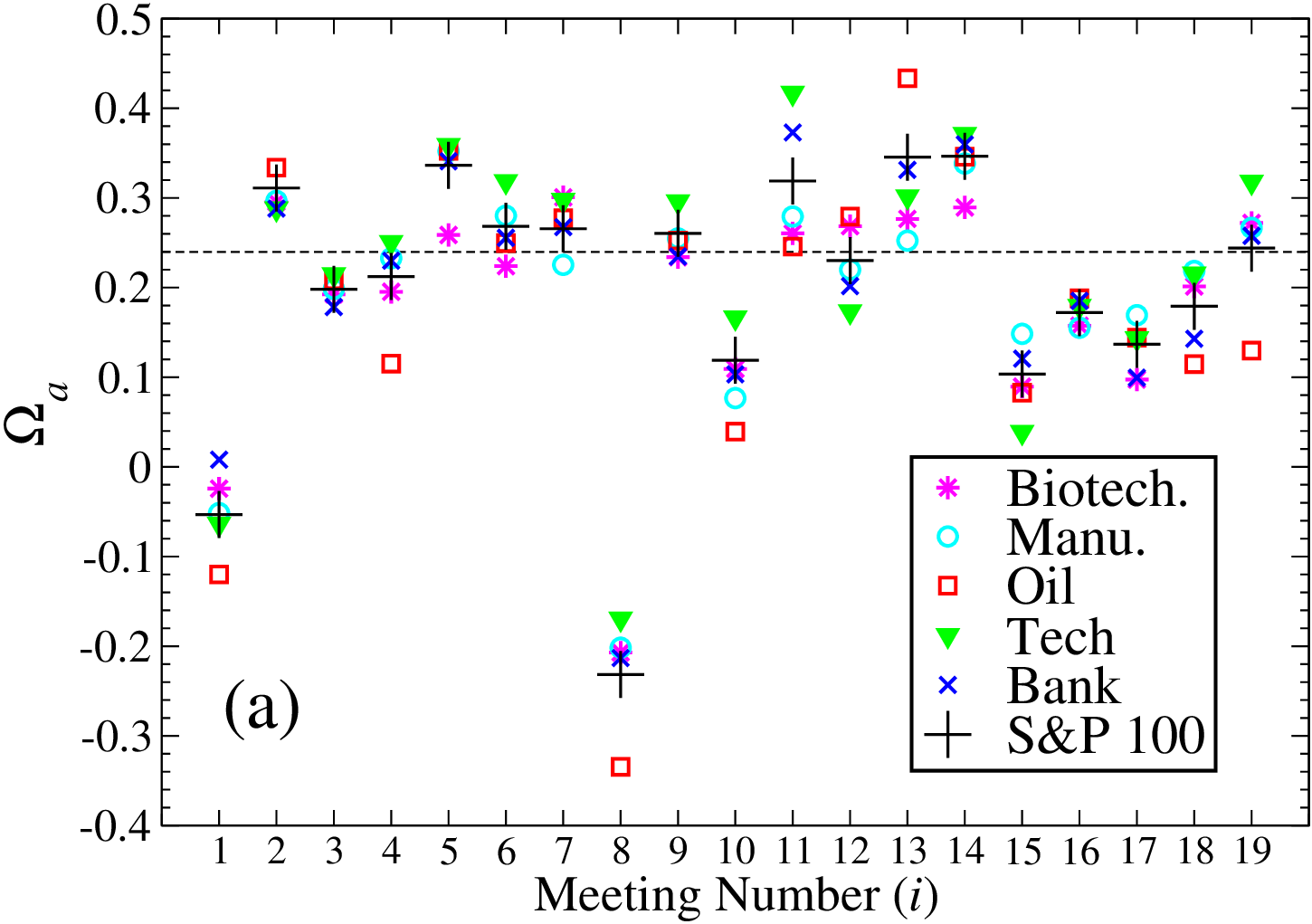}}\\
\centering{\includegraphics[width=0.7\textwidth]{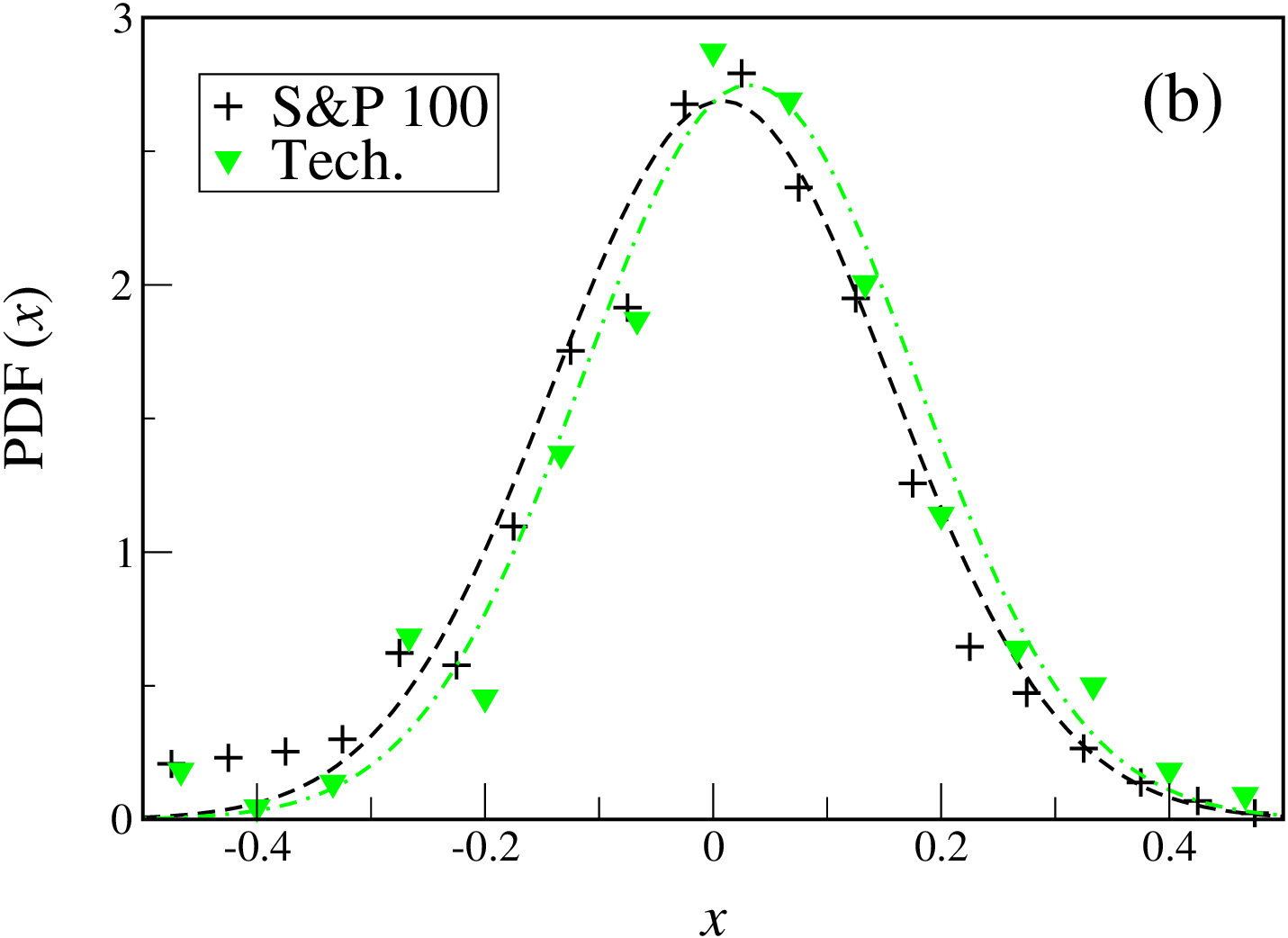}}
 \caption{    \label{SectorOmega} (Color online) (a) A comparison of $\Omega_{a}$ for 5 sectors with volatility
threshold $q=3$ suggests a broad universal market response to FOMC news. The Tech sector tends to have the largest
average $\Omega_{a}$, where large $\Omega$ values corresponds to faster relaxation. The horizontal straight line
represents the mean   $ \overline{\Omega}_{a} = 0.24 \pm 0.08$, averaged over all stocks in the S\&P100 and all
scheduled meetings (excluding the unscheduled meetings $i=\{1,4,8\}$). 
 (b)  Probability density function $P(x)$ of the variable   $x \equiv x_{a,i}^{j} = \Omega_{a,i}^{j}-
\langle\Omega_{a,i} \rangle$, which correspond to individual $\Omega_{a,i}^{j}$ values centered around the average
exponent $\langle \Omega_{a,i} \rangle$ of a given meeting $i$. We conclude from a Z-test at the $\alpha = 0.0005$
significance level that Tech sector Omori exponents are larger on average, $\langle x \rangle_{Tech} > \langle x
\rangle_{SP100}$. 
Hence, since larger $\Omega$ values correspond to shorter relaxation time, we find that the Tech sector stocks responds more
quickly to FOMC news, possibly as a result of relatively intense trading activity among these stocks. 
 }
\end{figure}

\begin{figure}[t]
\centering{\includegraphics[width=0.7\textwidth]{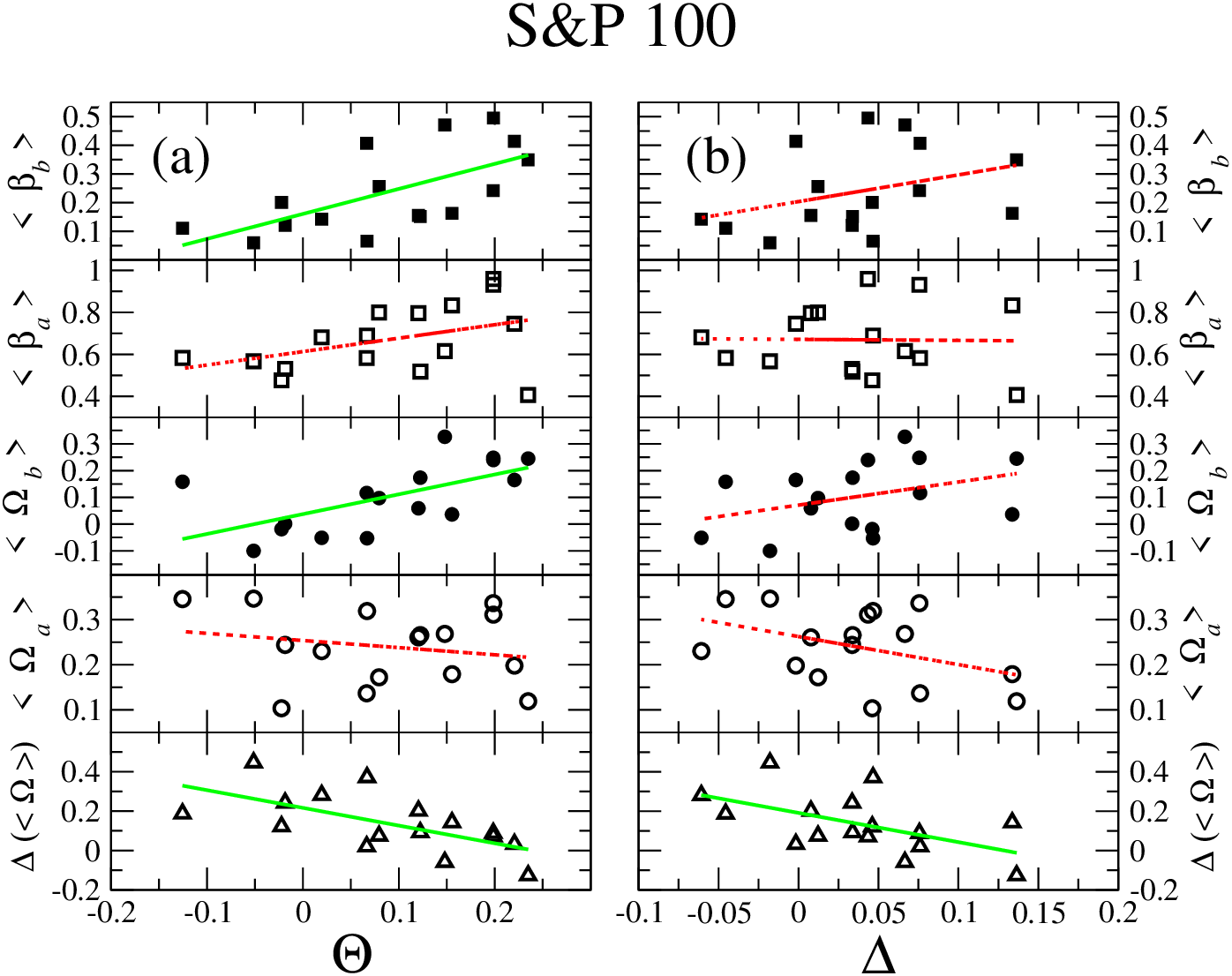}}\\

\centering{\includegraphics[width=0.7\textwidth]{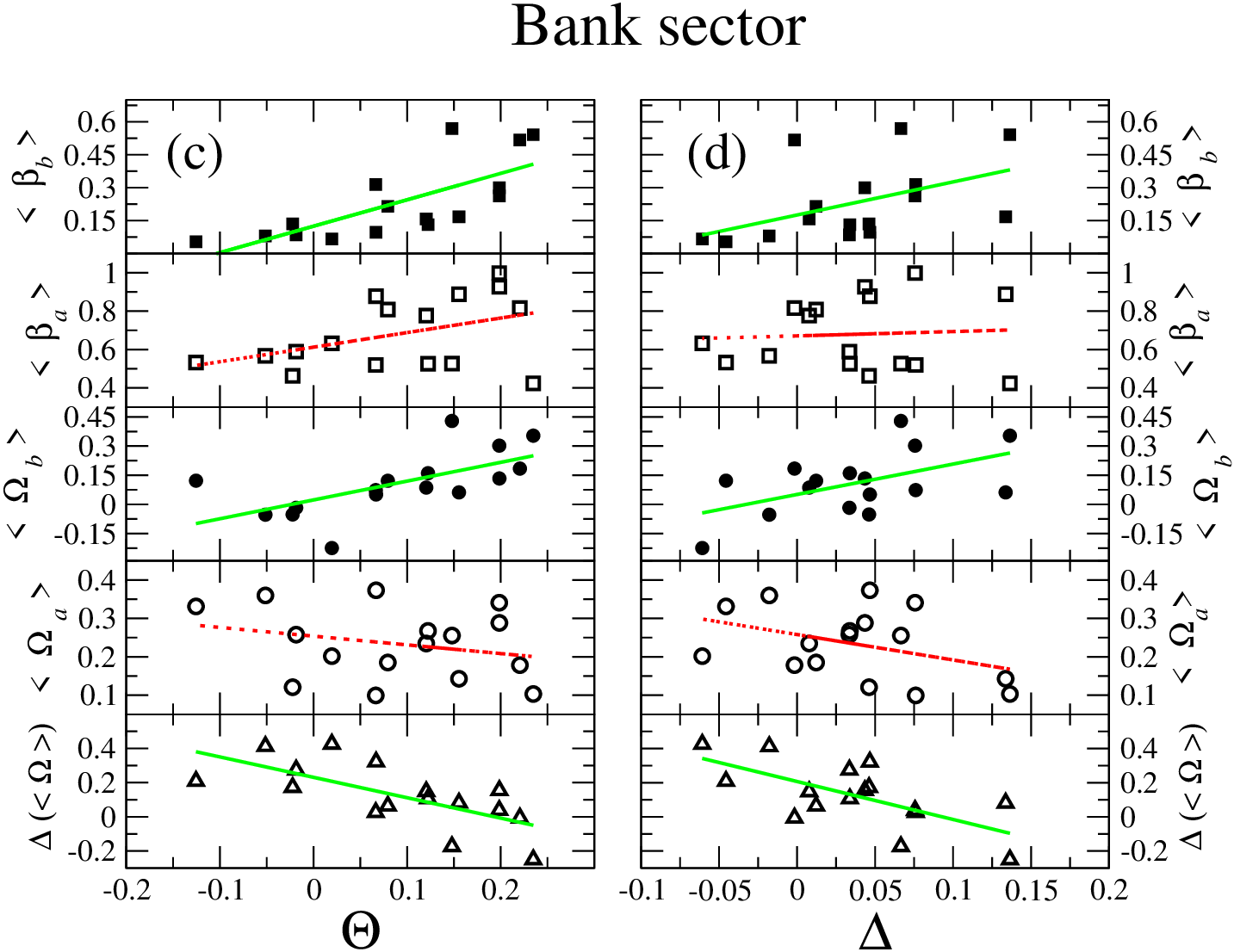}}
 \caption{
  \label{ThetaIntraday} (Color online) The relation between the size of the financial shock, quantified by the S\&P 100
volatility Omori law parameters $\langle \beta_{b}\rangle$, $\langle \beta_{a}\rangle$, $\langle \Omega_{b}\rangle$, 
$\langle  \Omega_{a} \rangle$, and $ \Delta(\langle \Omega \rangle) = \langle 
\Omega_{a} \rangle-\langle  \Omega_{b} \rangle$, and the size of the FOMC news, quantified through the metrics 
$\Theta$ representing market anticipation and  $\Delta$ representing market surprise. All trends are consistent with the
hypothesis that a strong anticipation of an interest rate change, and the element of surprise inherent in the FOMC decision,
result in a market perturbation that is significant in scale, and broad across the market. 
 Linear regressions of S\&P100 data (a,b) and bank sector data (c,d) are provided for visual aide. Linear regressions
that pass the ANOVA F-test (rejecting null hypothesis that regression slope $m=0$) at the $\alpha = 0.05$ significance
level are solid green; regressions that fail to pass the F-test at the $\alpha = 0.05$ significance level are
dashed red.}
\end{figure}

\end{widetext} 


\begin{thebibliography}{99}
\bibitem{ThetaDeltaPar} For the calculation of  $\Theta$ and $\Delta$, we choose values of $L$ and $\lambda$ to be on the order of a couple trading weeks prior to the
announcement, 
 so that we isolate fresh speculation leading into the meeting. The parameter $\lambda$ provides an effective cutoff
period, after which the weights begin
 to decrease quickly. Conversely, the weights corresponding to days close to the meeting, $\Delta t = 0$, are
effectively constant. The values of 
 $\Theta$ and $\Delta$ do not change much with varying choice of $L \in [5,15]$ or $\lambda \in [5,15]$. Without loss of
generality, we choose $L=15$ days, and $\lambda = 10$ days.
\bibitem{data} We find historical information about FOMC meetings using resources at the Federal Reserve websit and using newspaper archives.
Intraday time of announcement, $T$, are often quoted in New York Times finance articles by  Richard W. Stevenson the day
after FOMC announcements. They are also evident in the intraday Omori plots of $N(t)$ in Fig. 4.
\bibitem{powlawfit} In Fig. 4 we plot $N(t)$, where $t=0$ corresponds to the opening bell at 9:30 AM. Quantitatively, we calculate the Omori exponents 
using the displaced time $\tau$ from the announcement time $T$. 
In terms of the  displaced time $\tau \equiv \vert t- T_{i} \vert$, we
separate $N_{i}(t)$ into two separate time series $N_{b,i}(\tau)=N_{i}(T_{i})- N_{i}(\vert t- T_{i}\vert)$ where $p <
t < T_{i}$ \ and \
$N_{a,i}(\tau)=N_{i}(t-T_{i})- N_{i}(T_{i})$  with $t > T_{i}$.  The parameter $p$ is a ``padding''  that eliminates
the last 60 minutes of $N_{b,i}(\tau)$ (the first 60 minutes of $N_{i}(t)$), which eliminates opening effects and
improves the regression analysis around $T_{i}$.
\end{thebibliography}
\end{document}